
\newif\iffigs\figstrue

%
\let\useblackboard=\iftrue
%
%
\newfam\black

\input harvmac.tex

\input epsf

\newcount\figno
\figno=0
\def\fig#1#2#3{
\par\begingroup\parindent=0pt\leftskip=1cm\rightskip=1cm\parindent=0pt
\baselineskip=11pt
\global\advance\figno by 1
\midinsert
\epsfxsize=#3
\centerline{\epsfbox{#2}}
\vskip 12pt
{\bf Fig.\ \the\figno: } #1\par
\endinsert\endgroup\par
}
\def\figlabel#1{\xdef#1{\the\figno}}
\def\encadremath#1{\vbox{\hrule\hbox{\vrule\kern8pt\vbox{\kern8pt
\hbox{$\displaystyle #1$}\kern8pt}
\kern8pt\vrule}\hrule}}
\overfullrule=0pt

\def\Title#1#2{\rightline{#1}
\ifx\answ\bigans\nopagenumbers\pageno0\vskip1in%
\baselineskip 15pt plus 1pt minus 1pt
\else
\def\listrefs{\footatend\vskip 1in\immediate\closeout\rfile\writestoppt
\baselineskip=14pt\centerline{{\bf References}}\bigskip{\frenchspacing%
\parindent=20pt\escapechar=` \input
refs.tmp\vfill\eject}\nonfrenchspacing}
\pageno1\vskip.8in\fi \centerline{\titlefont #2}\vskip .5in}

\ifx\answ\bigans\def\tcbreak#1{}\else\def\tcbreak#1{\cr&{#1}}\fi
\useblackboard
\message{If you do not have msbm (blackboard bold) fonts,}
\message{change the option at the top of the tex file.}
\font\blackboard=msbm10 
\font\blackboards=msbm7
\font\blackboardss=msbm5
\textfont\black=\blackboard
\scriptfont\black=\blackboards
\scriptscriptfont\black=\blackboardss
\def\Bbb#1{{\fam\black\relax#1}}
\else
\def\Bbb#1{{\bf #1}}
\fi
%
\def\yboxit#1#2{\vbox{\hrule height #1 \hbox{\vrule width #1
\vbox{#2}\vrule width #1 }\hrule height #1 }}
\def\fillbox#1{\hbox to #1{\vbox to #1{\vfil}\hfil}}
\def\ybox{{\lower 1.3pt \yboxit{0.4pt}{\fillbox{8pt}}\hskip-0.2pt}}
\def\np#1#2#3{Nucl. Phys. {\bf B#1} (#2) #3}
\def\pl#1#2#3{Phys. Lett. {\bf #1B} (#2) #3}

\def\physrev#1#2#3{Phys. Rev. {\bf D#1} (#2) #3}

\def\comments#1{}

\def\half{{1\over 2}}

\def\bra#1{{\langle}#1|}
\def\ket#1{|#1\rangle}

\def\a{\alpha}

\def\II{\relax{I\kern-.07em I}}

\def\hk{{hyperk\"ahler}}

\def\IZ{\relax\ifmmode\mathchoice
{\hbox{\cmss Z\kern-.4em Z}}{\hbox{\cmss Z\kern-.4em Z}}
{\lower.9pt\hbox{\cmsss Z\kern-.4em Z}}
{\lower1.2pt\hbox{\cmsss Z\kern-.4em Z}}\else{\cmss Z\kern-.4em
Z}\fi}
\def\IB{\relax{\rm I\kern-.18em B}}
\def\IC{\bf C}
\def\ID{\relax{\rm I\kern-.18em D}}
\def\IE{\relax{\rm I\kern-.18em E}}
\def\IF{\relax{\rm I\kern-.18em F}}
\def\IG{\relax\hbox{$\inbar\kern-.3em{\rm G}$}}
\def\IGa{\relax\hbox{${\rm I}\kern-.18em\Gamma$}}
\def\IH{\relax{\rm I\kern-.18em H}}
\def\II{\relax{\rm I\kern-.18em I}}
\def\IK{\relax{\rm I\kern-.18em K}}
\def\IP{\relax{\rm I\kern-.18em P}}

\useblackboard
\def\IZ{\relax\Bbb{Z}}
\fi

\font\cmss=cmss10 \font\cmsss=cmss10 at 7pt
\def\IR{\relax{\rm I\kern-.18em R}}

\def\BR{\IR}
\def\BZ{\IZ}
\def\BR{\IR}
\def\BC{\IC}



\def\lim{{lim}}

\input epsf

\def\SUSY#1{{{\cal N}= {#1}}}                   
\def\dg{{\dagger}}
\def\wdg{{\wedge}}                              



\def\MR#1{{{\BR}^{#1}}}               
\def\MC#1{{{\BC}^{#1}}}               

\def\MR#1{{{\BR}^{#1}}}               
\def\MC#1{{{\BC}^{#1}}}               
\def\MS#1{{{\bf S}^{#1}}}               
\def\MT#1{{{\bf T}^{#1}}}               
\def\MHT#1{{{\bf \widetilde{T}}^{#1}}}               

\def\px#1{{\partial_{#1}}}              





\def\trp#1{{{\rm tr}\{ {#1} \} }}            

\def\rep#1{{{\bf {#1}}}}                      



\def\hepth#1{{\it hep-th/{#1}}}

\def\frac#1#2{{{{#1}}\over {{#2}}}}           

\def\u{{\mu}}
\def\v{{\nu}}
\def\b{{\beta}}

\def\lam{{\lambda}}



\def\Modsp{{\cal M}}     
\def\hK{{hyper-K\"ahler}}

\def\Vol#1{{{V\!\!ol\left({#1}\right)}}}    
\def\tw{{\theta}}         
\def\btw{{\eta}}         

\def\Ph{{{\Phi}}}

\def\jhep#1#2#3{{{{\bf JHEP {#1}}({#2}){#3} }}}




%
\Title{ \vbox{\baselineskip12pt\hbox{hep-th/9805045, PUPT-1787}}}
{\vbox{
\centerline{On the Twisted $(2,0)$ and Little-String Theories}}}
\centerline{
Yeuk-Kwan E. Cheung\footnote{$^1$}{cheung@viper.princeton.edu},
Ori J. Ganor\footnote{$^2$} {origa@puhep1.princeton.edu} and
Morten Krogh\footnote{$^3$}{krogh@phoenix.princeton.edu}
}
\smallskip
\smallskip
\centerline{Department of Physics}
\centerline{Jadwin Hall}
\centerline{Princeton University}
\centerline{Princeton, NJ 08544, USA}
\bigskip
\bigskip
\noindent
We study the compactification of the $(2,0)$ and type-II little-string
theories on $S^1$, $T^2$ and $T^3$ with an R-symmetry twist that
preserves half the supersymmetry.
We argue that it produces the same moduli spaces of vacua
as compactification of the
$(1,0)$ theory with $E_8$ Wilson lines given by a maximal embedding
of $SU(2)$. In certain limits, this
reproduces the moduli space of $SU(2)$ with a massive adjoint
hyper-multiplet. In the type-II little-string theory case,
we observe a peculiar phase transition where the strings
condense. We conjecture a generalization to more than two 5-branes 
which involves instantons on non-commutative $T^4$.
We conclude with open questions.

\Date{May, 1998}



\lref\rWAdSII{E. Witten, 
   {``Anti-de Sitter Space, Thermal Phase Transition, and Confinement
    in Gauge Theories,''} \hepth{9803131}}

\lref\rG{O.J. Ganor,
  {``Toroidal Compactification of Heterotic 6D  
  Non-Critical Strings Down to Four Dimensions,''}
  \np{488}(1997){223}, \hepth{9608109}}

\lref\rGMS{O.J. Ganor, D.R. Morrison and N. Seiberg,
  {``Branes, Calabi-Yau Spaces, and Toroidal Compactification
  of the $N=1$ Six-Dimensional $E_8$ Theory,''} 
  \np{487}{1997}{93}, \hepth{9610251}}

\lref\rIMS{K. Intriligator, D.R. Morrison and N. Seiberg,
  {``Five-Dimensional Supersymmetric Gauge Theories and Degenerations
  of Calabi-Yau Spaces,''} \np{497}{97}{56-100}, \hepth{9702198}}

\lref\rWitCOM{ E. Witten,
  {``Some Comments on String Dynamics,''}
  \hepth{9507121},
  published in {\it ``Future perspectives in string theory,''} 501-523.}

\lref\rStrOPN{A. Strominger,
  {``Open p-Branes,''} \pl{383}{1996}{44-47}, \hepth{9512059}}

\lref\rWitNGT{E. Witten,
  {``New ``Gauge'' Theories In Six Dimensions,''}
  \hepth{9710065}}

\lref\rSeiVBR{N. Seiberg,
  {``New Theories in Six-Dimensions and
  Matrix Description of M-theory on $T^5$ and $T^5/Z_2$,''}
  \hepth{9705221}, \pl{408}{97}{98}}

\lref\rBFSS{T. Banks, W. Fischler, S.H. Shenker and L. Susskind,
  {``M Theory As A Matrix Model: A Conjecture,''}
  \hepth{9610043}, \physrev{55}{1997}{5112-5128}}

\lref\rDVVQ{R. Dijkgraaf, E. Verlinde, H. Verlinde,
  {``BPS Quantization Of The 5-Brane,''}
  \np{486}{97}{77}, \hepth{9604055}}
\lref\rDVVS{R. Dijkgraaf, E. Verlinde, H. Verlinde,
  {``BPS Spectrum Of The 5-Brane And Black-Hole Entropy,''}
  \np{486}{97}{89}, \hepth{9604055}}

\lref\rSWSIXD{N. Seiberg and E. Witten,
  {``Comments On String Dynamics In Six-Dimensions,''}
  \hepth{9603003}, \np{471}{1996}{121}}

\lref\rKS{A. Kapustin and S. Sethi,
  {``The Higgs Branch of Impurity Theories,''} \hepth{9804027}}

\lref\rSeiSTN{N. Seiberg,
  {\it ``Notes on Theories with 16 Supercharges,''}
  \hepth{9705117}}

\lref\rKV{S. Kachru and C. Vafa,
  {``Exact Results For $N=2$ Compactifications
  Of Heterotic Strings,''} \np{450}{95}{69}, \hepth{9505105}}

\lref\rWitFBR{E. Witten,
  {``Solutions Of Four-Dimensional Field Theories Via M Theory,''}
  \np{500}{1997}{3--42},\hepth{9703166}}

\lref\rBDS{T. Banks, M.R. Douglas and N. Seiberg,
  {``Probing F-theory With Branes,''}
  \pl{387}{1996}{278--281}, \hepth{9605199}}

\lref\rSenFO{A. Sen,
  {``F-theory and Orientifolds,''}
  \np{475}{1996}{562-578}, \hepth{9605150}}

\lref\rSeiIRD{N. Seiberg,
  {``IR Dynamics on Branes and Space-Time Geometry,''}
  \pl{384}{1996}{81--85}, \hepth{9606017}}

\lref\rSav{S. Sethi,
  {``The Matrix Formulation of Type IIB Five-Branes,''}
  \hepth{9710005}}

\lref\rGS{O.J. Ganor and S. Sethi,
  {``New Perspectives on Yang-Mills Theories With Sixteen Supersymmetries,''}
  \hepth{9712071}}

\lref\rGK{S.S. Gubser and I.R. Klebanov,
  {``Absorption by Branes and Schwinger Terms in the World Volume Theory,''}
  \pl{413}{1997}{41--48}, \hepth{9708005}}

\lref\rMS{J. Maldacena and A. Strominger,
  {``Semiclassical Decay of Near Extremal 5-Branes,''}
  \jhep{12}{97}{008}, \hepth{9710014}}

\lref\rSWGDC{N. Seiberg and E. Witten,
  {``Gauge Dynamics And Compactification To Three Dimensions,''}
  \hepth{9607163}}

\lref\rSeiHOL{N. Seiberg, {``Naturalness Versus Supersymmetric 
  Non-Renormalization Theorems,''}
 \pl{318}{1993}{469--475}, {\tt hep-ph/9309335}.}

\lref\rAspin{P. Aspinwall,
  {``K3 Surfaces and String Duality,''} \hepth{9611137}} 

\lref\rSeiFIV{N. Seiberg,
  {``Five Dimensional SUSY Field Theories, Non-trivial Fixed Points
  and String Dynamics,''} \pl{388}{1996}{753-760},\hepth{9608111}}

\lref\rSWII{N. Seiberg and E. Witten,
  {``Monopoles, Duality and Chiral Symmetry Breaking in $N=2$
  Supersymmetric QCD,''} \np{431}{1994}{484--550}, \hepth{9408099}}

\lref\rBarak{B. Kol,
  {``On 6d ``Gauge'' Theories with Irrational Theta Angle''},
  \hepth{9711017}}

\lref\rDH{M.R. Douglas and C. Hull,
  {``D-branes and the Noncommutative Torus,''}
  \hepth{9711165}}

\lref\rSavLen{S. Sethi and L. Susskind,
  {``Rotational Invariance in the M(atrix) Formulation
  of Type IIB Theory,''} \pl{400}{1997}{265--268}, \hepth{9702101}}

\lref\rTomNat{T. Banks and N. Seiberg,
  {``Strings from Matrices,''} \np{497}{1997}{41--55}, \hepth{9702187}}

\lref\rPolWit{J. Polchinski and E. Witten,
  {``Evidence For Heterotic - Type I String Duality,''}
  \np{460}{1996}{525}, \hepth{9510169}}

\lref\rBerDou{M. Berkooz and M.R. Douglas,
  {``Five-branes in M(atrix) Theory,''} \hepth{9610236}}

\lref\rSeiWHY{N. Seiberg,
  {``Why is the Matrix Model Correct?,''} 
  \physrev{79}{1997}{3577--3580}, \hepth{9710009}}

\lref\rSenTD{A. Sen,
  {``D0 Branes on $T^n$ and Matrix Theory,''}
  \hepth{9709220}}

\lref\rWati{W. Taylor,
  {``D-brane field theory on compact spaces,''}
  \hepth{9611042}, \pl{394}{1997}{283}.}

\lref\rCDS{A. Connes, M.R. Douglas and A. Schwarz,
  {``Noncommutative Geometry and Matrix Theory: Compactification on Tori,''}
  \hepth{9711162}}

\lref\rCK{Y.-K. E. Cheung and M. Krogh,
  {``Noncommutative Geometry from 0-branes in a Background B-field,''}
  \hepth{9803031}}

\lref\rProg{Work in progress.}

\lref\rNS{N.Nekrasov, A.Schwarz,
  {``Instantons on noncommutative $R^4$ and (2,0) superconformal
  six dimensional theory,''} \hepth{9802068}}

\lref\rMicha{M. Berkooz,
  {``Non-local Field Theories and the Non-commutative Torus,''}
  \hepth{9802069}}

\lref\rGTest{O.J. Ganor,
  {``A Test Of The Chiral E8 Current Algebra On A 6D
  Non-Critical String,''} \np{479}{1996}{197--217}, \hepth{9607020}}

\lref\rKMV{A. Klemm, P. Mayr and C. Vafa, 
  {``BPS States of Exceptional Non-Critical Strings,''} \hepth{9607139}}  

\lref\rMNWI{J.A. Minahan, D. Nemeschansky and N.P. Warner,
  {``Investigating the BPS Spectrum of Non-Critical $E_n$ Strings,''}
  \np{508}{1997}{64--106}, \hepth{9705237}} 

\lref\rMNWII{J. A. Minahan, D. Nemeschansky and N. P. Warner,
  {``Partition Functions for BPS States of the Non-Critical $E_8$ String,''}
  Adv.Theor.Math.Phys. 1 (1998) 167-183, \hepth{9707149}}

\lref\rMNVW{J. A. Minahan, D. Nemeschansky, C. Vafa and N. P. Warner,
  {``E-Strings and $N=4$ Topological Yang-Mills Theories,''}
  \hepth{9802168}}


\lref\rKumVaf{A. Kumar and C. Vafa,
  {``U-Manifolds,''} \pl{396}{1997}{85-90}, \hepth{9611007}} 


\newsec{Introduction}

In this paper we will study the compactification of the $(2,0)$
theory and the little-string theory on $\MS{1}$, $\MT{2}$ and $\MT{3}$.
The $(2,0)$-theory describes the low-energy modes coming from type-IIB
on an $A_{k-1}$ singularity \rWitCOM\ or, equivalently,
 $k$ 5-branes of M-theory \rStrOPN.
The little-string theory is the theory of $k$ type-II NS5-branes
decoupled from gravity \rSeiVBR.
In order to get an interesting low-energy question we will
twist the boundary conditions along $\MT{d}$ by elements of
the $Spin(5)$ (or $Spin(4)$ for the little-string theory) R-symmetry.
In this way we obtain new kinds of theories with 8 supersymmetries.
The aim of this paper is to find the low-energy description of
these theories. We will present an explicit construction in the
case $k=2$ and suggest a conjecture for higher $k$.
The construction for $k=2$ involves the moduli space of the
heterotic 5-brane wrapped on tori.
The conjecture for any $k$
involves moduli spaces of instantons on non-commutative tori.
In certain limits we recover the known moduli spaces of
Super-Yang-Mills theories  with a massive adjoint hypermultiplet.
In the compactified little-string theories, examination of the moduli
space shows that for certain values of the external parameters 
there is a phase transition to a phase where little-strings condense.

The paper is organized as follows.
In section (2) we explain our notation, present the problem
and discuss the parameters on which the compactifications depend.
In section (3) we present the general solution for $k=2$.
In section (4) we study in more detail
various limiting cases of the solution. In particular, we
study the limits where Super-Yang-Mills is obtained.
In subsection (4.2) we observe the phase transition.
In section (5) we explain the relation between the twist and
the mass of the adjoint hypermultiplets in the effective low-energy
description of Super-Yang-Mills.
In section (6) we discuss in more detail what it means to 
twist the little-string theories. We study what happens to the
twists after T-duality and suggest that the R-symmetry twists
are a special case of a more general twist.
In section (7) we present the conjecture for higher $k$ and
the relation to moduli spaces of instantons on non-commutative
tori.
In section (8) we briefly discuss the questions raised
in the M(atrix)-approach to these twists.
We end with a discussion and open problems.


\newsec{The problem}
The problem that we are going to study is to find
the Seiberg-Witten curves of certain $\SUSY{2}$ theories in 3+1D
and to find the \hk\ moduli space of certain
$\SUSY{4}$ theories in 2+1D.
The $\SUSY{2}$ theories will be obtained by compactifying the
$(2,0)$ theory or, slightly generalizing,  the little-string theory,
on $\MT{2}$ with twisted R-symmetry
boundary conditions along the sides of
the torus.
The $\SUSY{4}$ theories in 2+1D are similarly obtained by
compactification on $\MT{3}$.
In this section we will describe the setting and the notation.

\subsec{Definitions}
Let us denote by $T(k)$ the $(2,0)$ low-energy theory of $k$ 5-branes
of M-theory \refs{\rWitCOM,\rStrOPN}.
We denote 
by $S_A(k)$ ($S_B(k)$) the theory of $k$ type-IIA (type-IIB)
NS5-branes in the limit
when the string coupling goes to zero keeping the string
tension fixed \rSeiVBR.
Compactified on a circle, these two theories are T-dual.
$T(k)$ is often called ``{\it the $(2,0)$ theory}'' and $S(k)$
is referred to as ``{\it the little-string theory}''.

\subsec{The $(2,0)$ theory and the little-string theories}
When $T(k)$ is compactified on $\MT{2}$ we get a 3+1D theory
which at low-energy becomes $k$ free vector multiplets (at
generic points in the moduli space).
The vector-multiplet moduli space is $(\MS{1}\times\MR{5})^k/S_k$
where the size of $\MS{1}$ is $A^{-1/2}$
and $A$ is the area of $\MT{2}$.
When we compactify $T(k)$ on $\MT{3}$ the low-energy
is (generically on the moduli space) given by a $\sigma$-model
on the \hk\ manifold $(\MT{3}\times\MR{5})^k/S_k$.
The $\MT{3}$ in the moduli space has the same
shape as the physical $\MT{3}$ but its volume is $V^{-1/2}$,
where $V$ is the volume of the physical $\MT{3}$.
(See \rSeiSTN\ for review.)
$S_A(k)$ has a low-energy
description given by 5+1D SYM and has a scale $M_s$. The scale
is related to the SYM coupling constant $M_s^{-1}$.
The parameters of the compactification
are now the metric on $\MT{3}$ and also the NSNS 2-form
on $\MT{3}$. The 2-form couples as a $\theta$-angle in the
effective 5+1D low-energy SYM, i.e.  as $\int B\wdg \trp{F\wdg F}$.
Together they parameterize 
\eqn\slfour{
SO(3,3,\BZ)\backslash SO(3,3,\BR)/(SO(3)\times SO(3))
 = SL(4,\BZ)\backslash SL(4,\BR)/ SO(4).
}
The moduli space is given by
$(\MT{4}\times\MR{4})^k/S_k$ where $\MT{4}$ has the shape
given by the point in $SL(4,\BZ)\backslash SL(4,\BR)/ SO(4)$
and has a fixed volume $M_s^2$.

We have to mention that the arguments of \rMS\  (see also \rGK)
show that the theories $S(k)$ are far more complicated
than the $T(k)$ theories, in the sense that they have a continuous
spectrum starting at energy around $M_s$ and this spectrum describes
graviton states propagating in a weakly coupled throat.
Below the scale $M_s$ there is a discrete spectrum (up to
the  effect of the $4k$ non-compact scalars).
Since there is a mass gap, one can still ask low-energy
questions, as we are doing.

\subsec{R-symmetry Wilson lines}
The compactifications discussed above have 16 supersymmetries
and therefore the moduli spaces obtained in 2+1D are flat
and only their global structure is interesting.
To get interesting metrics on the moduli space we need to break
the supersymmetry down by ${1\over 2}$.
This can be done as follows.
Suppose we identify a global symmetry of the $(2,0)$ theory.
When we compactify on $\MS{1}$ of radius $R$ and coordinate
$0\le x\le 2\pi R$, we can glue
the points $x=0$ and $x=2\pi R$ by adding a twist of the global
symmetry. When we compactify on $\MT{3}$ we can twist along all
3 directions so long as the twists commute. The global symmetry
of $T(k)$ is the $Spin(5)$ R-symmetry. Such a twist
has been recently used in \rWAdSII\ to break the supersymmetry
of the $(2,0)$ theory in compactifications.

When we compactify the little-string theory $S_A(k)$ ($S_B(k)$) it is not
immediately obvious that we can use such a twist because
the space-time interpretation is not unique. However, since
we can embed the twist as a geometrical twist in type-IIA,
the question is well defined. We will elaborate more on that point
in section (6).

Let us now take the $(2,0)$ theory $T(k)$ on $\MT{3}$ with
three commuting twists $g_1,g_2,g_3\in Spin(5)$ along $\MT{3}$.
The 16 super-charges of $T(k)$ transform as a space-time spinor
which also has indices in the $\rep{4}$ of $Spin(5)$.
The condition that 8 supersymmetries will be preserved
is the condition that $g_1,g_2,g_3$ preserve a two-dimensional
subspace of the representation $\rep{4}$
of $Spin(5)$. This becomes the following condition.
Take $SU(2)_B\times SU(2)_U = Spin(4)\subset Spin(5)$
and let $g_1,g_2,g_3$ be 3 commuting elements in the first
$SU(2)_B$ factor. This is the generic twist which preserves
$\SUSY{4}$ in 2+1D.
Similarly, for the little-string theory $S(k)$ the R-symmetry
is $Spin(4)$ and we need,
$$
g_1,g_2,g_3 \in SU(2)_B\subset SU(2)_B\times SU(2)_U = Spin(4).
$$
Since the $g_i$'s are commuting they can be taken inside
a $U(1)$ subgroup of $SU(2)_B$.
Then $g_i = e^{i\tw_i}\in U(1)\subset SU(2)_B$.
The subscripts $B$ and $U$ are short for ``broken''
and ``unbroken'' respectively.
We can now ask what is the low-energy description of $T(k)$,
$S_A(k)$ and $S_B(k)$ compactified, in turn,
 on $\MS{1}$, $\MT{2}$ and $\MT{3}$ with twists $\tw_i$.
The most general question is about $S(k)$ on $\MT{3}$ since
all others can be obtained by taking appropriate limits.
The low-energy description in 2+1D is a $\sigma$-model on a
$4(k-1)$-dimensional \hk\ manifold. We will always ignore the
decoupled ``center of mass''.
Furthermore, as will be elaborated in section (4),
in appropriate limits we obtain
3+1D or 2+1D $SU(k)$ SYM with a massive adjoint hypermultiplet.

What is the external parameter space?
The parameter space for the metric and $B$ fields on $\MT{3}$
is given by \slfour. The parameter space for conjugacy
classes of three commuting $SU(2)$ R-symmetry twists along
$\MT{3}$ is given by $\MHT{3}/\BZ_2$ where $\MHT{3}$ is 
the torus dual to $\MT{3}$
and $\BZ_2$ is the Weyl group of $SU(2)$.
However, with R-symmetry twists,
we can no longer divide by the full T-duality group
$SO(3,3,\BZ)$ (see the discussion in section (6)).
This means that the parameter space is a fibration of
$(\MHT{3}/\BZ_2)$ over 
$$
SL(3,\BZ)\backslash SO(3,3,\BR)/(SO(3)\times SO(3)).
$$

\subsec{Why is the problem not trivially solved by M-theory?}
Let us explain why we cannot just read off the
SW-curves and moduli spaces from M-theory.
To be specific, let us take the 6-dimensional non-compact space defined 
 as an $\MR{4}$-fibration over $\MT{2}$ with $Spin(4)$ twists
along the cycles of the $\MT{2}$. This is the geometric realization
of the R-symmetry twist, that we mentioned above (see section
(6) for a more detailed discussion).
M-theory compactified on this space preserves 16 supersymmetries if
the two twists $\tw_1,\tw_2$ are taken inside
$SU(2)_B\subset SU(2)_B\times SU(2)_U = Spin(4)$.
Let us wrap $k$ 5-branes on $\MT{2}$. Given the success of the method
described in \rWitFBR\ one may at first sight wonder whether
the classical moduli space of the $k$ 5-branes immediately gives the right
answer. The answer is negative. There is, in fact, a big difference
between the situation in \rWitFBR\ and ours.
The construction of \rWitFBR\ was used to solve certain QCD questions.
As explained there,
QCD is {\it not} the low-energy description of 5-branes in M-theory.
It is not even an approximate one. QCD is only a good approximation
in the region of moduli space where the 5-branes are close together
and the 11$^{th}$ dimension is very small. When this parameter was increased
the dynamics of the system is completely changed except for
the vacuum states (i.e. the moduli of the vector-multiplets).
This relied on the fact that the parameter that deforms the system
from close NS5-branes and D4-branes in type-IIA to M5-branes
decoupled from the vector-multiplet moduli space (similarly to the decoupling
in \rKV\ and \refs{\rBDS,\rSeiIRD}).
The classical result was correct for the M5-brane limit because 
all the relevant sizes were much larger than $M_{Pl}$ (the Planck scale).

In our case, not all the relevant sizes of the M5-brane configuration
are large. Let $A$ be the area of $\MT{2}$ and let $\Phi$
be the modulus of the tensor multiplet in 5+1D. $\Phi$ is related
to the separation $y$ between the 5-branes as $\Phi\sim M_p^3 y$.
The interesting region in moduli space is $\Phi A \sim 1$.
This region is $M_p^3 A y \sim 1$ and at least one of $y$ or $A$
cannot be made large.


\newsec{Solution}
In this section we will consider the theory $S_A(2)$ compactified on $\MT{3}$. 
We recall that $S_A(2)$ is the theory living on 2 coincident NS 5-branes in 
type IIA in the limit of vanishing string coupling with string scale,
$M_s$, kept fixed. The compactified theory has a moduli space of vacua 
which is a \hk\ manifold. The purpose of this section is to find this 
\hk\ manifold as a function of the parameters of the compactification. 
These parameters are described above. There is the IIA string scale,
$M_s$ (which is already a parameter in 6 dimensions).
There is the metric, $G^A_{ij}$ 
and NS-NS 2-form, $B^A_{ij}$, on the $\MT{3}$. Here A denotes the
underlying type IIA theory. Finally, there are the 
holonomies of the $Spin(4)$ R-symmetry around the 3 circles. The holonomies 
are taken inside an $SU(2)_B$ subgroup of $Spin(4)$ to preserve half of the 
supersymmetries. The 3 holonomies must commute and can thus be taken 
inside $U(1) \subset SU(2)_B$. We denoted the holonomies $e^{i\tw_1},
e^{i\tw_2},e^{i\tw_3}$, where $\tw_i$ is periodic with period 
 $2\pi$. Furthermore the Weyl group of $SU(2)_B$ relates $\tw_i$ to 
$-\tw_i$. These are the parameters of the theory.

The moduli space of vacua has real dimension 4, since we are dealing 
with 2 5-branes and we throw away the center of mass motion. We want 
to find the metric on this as a function of $M_s, G^A_{ij},B^A_{ij}$
and $\tw_i$.
Our strategy will be to start at the special point $\tw_i =0$ 
and then later understand how to do the general case. At $\tw_i =0$  
the theory actually has $\SUSY{8}$ supersymmetry in 3 dimensions
(like $\SUSY{4}$ in 4 dimensions). Here the moduli space is just the 
classical one. At the origin of the moduli space the low energy theory 
is an $SU(2)$, $\SUSY{8}$ theory. There are also heavy Kaluza-Klein modes 
with masses that go like multiples of $1 \over R_i$, where $R_i$ are 
the radii of the circles. In $\SUSY{4}$ language the multiplet is 
a vector-multiplet and an adjoint hypermultiplet. On the the moduli 
space of vacua $SU(2)$ is broken to $U(1)$. Dualizing the photon gives an 
extra scalar, so the vector-multiplet has 4 scalars. In the $\SUSY{8}$ 
theory the moduli space of vacua is 8 dimensional.
Four of the directions come 
from scalars in the hypermultiplet. These are lifted as soon as $\tw_i 
\neq 0$, because $\tw_i$ supply a mass to the hypermultiplet. We are 
really only interested in the 4 directions coming from scalars in the 
vector-multiplet. These 4 scalars are all compact. From the 5-brane point of 
view these scalars come about as follows. One of them is the relative 
position of the 5-branes on the 11$^{th}$ circle. The other 3 are the 
2-form living on the 5-brane with indices along the $\MT{3}$. These 4
 scalars are obviously compact. The Weyl group of the SU(2) gauge group 
changes the sign of all these. We thus see that the moduli space of 
vacua is $\MT{4} /\BZ_2$.
When we deform to $\tw_i\ne 0$, the moduli space remains compact.
The only compact 4 dimensional \hk\ manifolds are K3 and $\MT{4}$.
$\MT{4}/\BZ_2$ is topologically a K3 manifold with a 
singular metric. We thus conclude that for all parameters $G^A_{ij},B^A_{ij}, 
\tw_i$ the moduli space is topologically K3. We just need to find the 
\hk\ metric as a function of these parameters. 

Let us first recall the moduli space of \hk\ metrics on K3.
It is \rAspin,
$$
O(3,19,\BZ)\backslash O(3,19,\BR)/((O(3)\times O(19)) \times \MR{+}.
$$
$\MR{+}$ parameterizes the volume. This moduli space nicely 
coincides with the moduli space for Heterotic string theory on $\MT{3}$. 
This is a well-known consequence of the duality of M-theory on
K3 with heterotic on $\MT{3}$.
On the heterotic side the 
$\MR{+}$ denotes the dilaton. The space $O(3,19,\BR)/O(3) \times O(19)$ can 
be parameterized by the metric and NS-NS 2-form on the $\MT{3}$,
$G^H_{ij},B^H_{ij}$ and the Wilson lines around the 3 circles 
$V_1,V_2,V_3$. We will work with the $E_8 \times E_8$ Heterotic 
theory.
The reason for that will become clear in a moment.
There is a very nice way of obtaining the K3 on the M-theory side as 
a moduli space of vacua for a 3-dimensional $\SUSY{4}$ theory. 
This is the membrane of M-theory imbedded in $R^{1,6} \times 
K3$ with the world-volume along $R^{1,6}$ and at a point in K3. 
On the dual Heterotic side it corresponds to the 5-brane wrapped 
on $\MT{3}$ \rSeiIRD. This is thus the moduli space of the
$(1,0)$ little-string theory obtained from an NS5-brane
in the heterotic string by taking the coupling constant to zero \rSeiVBR.

Our aim can now be formulated as finding $G^H_{ij},B^H_{ij},V_1,V_2,V_3$ 
for given $G^A_{ij},B^A_{ij},\tw_i$.
According to the arguments of \rSeiHOL, the external parameters
can be combined into scalar components of auxiliary
vector-multiplets which are non-dynamical.
Supersymmetry then requires that the periods of the three 2-forms
which determine the \hk\ metric on the moduli space are linear
in these combinations of external parameters \rSWGDC.
To find the map subject to this restriction, we first examine $\tw_i=0$. 
We saw earlier that this was the $\SUSY{8}$ theory and the moduli 
space is $\MT{4} /\BZ_2$. Therefore, we can find the data of the $\MT{4}$ by 
classical analysis, starting from the $(1,0)$ tensor-multiplet 
living on the IIA 5-brane. (We have ignored the VEVs along 
the $(1,0)$ hypermultiplet direction.) The $(1,0)$ tensor-multiplet
is also the low-energy description of 
the $E_8 \times E_8$ Heterotic 5-brane and the scalar is 
compact since it corresponds to motion in the 11$^{th}$ direction, which 
is an interval. Let us compactify this theory on $\MT{3}$ with data 
$G^H_{ij},B^H_{ij},V_1,V_2,V_3$. To obtain the same moduli space of 
vacua as in the $S_A(2)$ case we need to set $G^H=G^A$,
$B^H=B^A$. What about $V_1,V_2,V_3$? The $S_A(2)$ theory had a $\MT{4} /\BZ_2$ 
as moduli space. $\MT{4} /\BZ_2$ has 16 $A_1$ singularities. This means 
that M-theory on this K3 has $SU(2)^{16}$ gauge symmetry. To achieve 
this we need very special Wilson lines. We can take $V_1$ to break 
$E_8 \times E_8$ to $SO(16) \times SO(16)$ and $V_2$ to break 
each SO(16) to $SO(8) \times SO(8)$ and $V_3$ to break each $SO(8)$
to $SO(4) \times SO(4)$. The unbroken symmetry group is thus 
$SO(4)^8 = SU(2)^{16}$ as desired.
These Wilson lines are unique up to $E_8\times E_8$ conjugation.
We can write down $V_1,V_2,V_3$ explicitly. The two $E_8$'s are treated 
symmetrically, so we restrict to one of them. Consider 
$\Gamma^8 \subset R^8$ where $\Gamma^8$ is the weight lattice of 
$E_8$. Recall that $\Gamma_8$ can be characterized as all sets 
$(a_1,\ldots,a_8)$ such that either all $a_i$ are half-integers
or all $a_i$ are integers. Furthermore $\sum a_i$ is even. A Wilson 
line around a circle can be specified by an element $V \in \MR{16}$ 
such that a ``state'' given by a weight vector $a$ is transformed as 
$e^{ia \cdot v}$\ on traversing the circle. In this notation
\eqn\wilsonlines{\eqalign{
V_1 &= (0,0, 0,0,0,0,0,1) \cr
V_2 &= (0,0,0,0,\half,\half,\half,\half) \cr
V_3 &= (0,0, \half,\half,0,0,\half,\half).
}}
Now we have the map in the case $\tw_i =0$. We will  make a  
proposal for the general case presently. The 16 singularities in $\MT{4} /\BZ_2$
 are due to an adjoint hypermultiplet becoming massless. When 
$\tw_i \neq 0$ the hypermultiplet is massive and we expect the 
singularities to disappear. Near the original singularities the 
theory now looks like pure $SU(2)$ SYM. This does not have any 
singularities \refs{\rSeiIRD,\rSWGDC}.
We thus see that the Wilson lines must change 
when we turn on $\tw_i$. We now make the following proposal. 
For nonzero $\tw_i$ we still have $G^H_{ij}=G^A_{ij}$,
$B^H_{ij}=B^A_{ij}$. The Wilson lines $W_1,W_2,W_3$ are taken to be,
$$
W_i = V_i + {\tw_i \over \pi}(\half,0,\half, 0,\half,0,\half,0),
$$
in the notation from above. This is the same as embedding 
$e^{\half i\tw_i}$ in the diagonal $SU(2) \subset SU(2)^{16} 
\subset E_8 \times E_8$. The coefficients of $\tw_i$ are 
chosen such that the period is $\tw_i \rightarrow \tw_i + 2\pi$. 

We do not have a proof of this proposal, but this certainly
satisfies the requirements of linearity in external
parameters, because this is also the moduli space of the compactified
$(1,0)$ little-string theory.
In the coming sections we will show that our proposal is consistent 
with string theory and field theory expectation.  

There is another very similar theory. This is the theory on 
2 coincident type IIB NS 5-branes in the limit of vanishing 
string coupling and fixed string mass. We call this theory 
$S_B(2)$. As soon as we compactify it on a circle it is T-dual 
to the theory studied above. When we compactify it on a $\MT{3}$ 
with R-symmetry twists we get a 3-dimensional theory with a 
K3 as the moduli space of vacua. Arguing exactly as in the IIA 
case we propose that this K3 is given in the same way as in 
the IIA case, except that we replace Heterotic $E_8 \times E_8$ 
with Heterotic SO(32). This is because the low energy description 
of the theory living on a IIB 5-brane is a gauge theory. The Heterotic 
SO(32) 5-brane is also described, at low energy, by a gauge theory. 
When we do the comparison at the point without an R-symmetry twist, 
the $\SUSY{8}$ point, the moduli spaces will automatically agree. 
This is analogous to the $\SUSY{8}$ point in the IIA case where 
we compared two tensor-multiplets. The T-duality between the IIA 
and IIB 5-brane theories on $\MT{3}$ fits very nicely with the 
T-duality between Heterotic SO(32) and Heterotic $E_8 \times 
E_8$ on $\MT{3}$ at the point $\tw_i=0$.
For $\tw_i\neq 0$, the R-symmetry twists do not remain R-symmetry
twists after T-duality.



\newsec{Limits}

Now that we have identified the moduli space of vacua for $S_A(2)$ 
compactified on $\MT{3}$ with arbitrary R-symmetry twists,
we can decompactify one or two of the circles to obtain 
the moduli space of vacua for $S_A(2)$ compactified to 
4 and 5 dimensions. Another limit is to take $M_s \rightarrow \infty$
in the $S_A(2)$ theory. This takes us to the $(2,0)$ theory, 
which we call $T(2)$. In this section we will consider these 
limits.

\subsec{Decompactification limits}

Let us first recall the correspondence between M-theory on K3 
and Heterotic $E_8 \times E_8$ on $\MT{3}$. M-theory on K3 has 
a Planck mass, $M_{Pl}$, and a moduli space 
$$
O(3,19,\BZ)\backslash O(3,19,\BR)/((O(3)\times O(19)) \times \MR{+}
$$
$\MR{+}$ denotes the volume of K3, $\Vol{K3}$. In Heterotic 
$E_8 \times E_8$ on $\MT{3}$ there is a string mass, $M_s$, and a 
moduli space, which is the same as for M-theory on K3. There 
is a 10-dimensional string coupling, $\lambda$. The $\MT{3}$ has 
a volume, $\Vol{\MT{3}}$, which is part of $O(3,19,\BR)/(O(3) \times O(19))$.
 Under the duality an M5-brane wrapped on K3 is mapped to the 
Heterotic string. Equating the tensions gives,
\eqn\volumen{
{M_{Pl}}^6\, \Vol{K3} = M_s^2.
}
Equating the 7-dimensional gravitational couplings gives,
\eqn\volum{
{M_{Pl}}^9\, \Vol{K3} = {M_s^8\Vol{\MT{3}} \over \lambda^2} 
}
We thus see, that the $\MR{+}$ on the Heterotic side is ${{\Vol{\MT{3}}}
\over {\lambda^2}}$, which of course is T-duality invariant. 
Eq.\volumen\ agrees with the fact that the volume of the moduli 
space of vacua of the Heterotic 5-brane is $M_s^2$ and 
${M_{Pl}}^6\,\Vol{K3}$ is the volume of the moduli space of the M 2-brane 
probe. We remember that scalar fields have dimension $\half$ in 
3 dimensions. A concrete way of tracing the duality between 
these two theories is to use T-duality from Heterotic $E_8 
\times E_8$ on $\MT{3}$ to Heterotic SO(32) on $\MT{3}$, and then 
S-duality to type-I on $\MT{3}$, then T-duality to type-IA on 
$\MT{3}$ which can be viewed as M-theory on K3.

Let us now consider the decompactification to 4 dimensions. 
This can be done by taking $\tw_3 =0$ and $R_3 \rightarrow 
\infty$. In this limit the K3 becomes elliptically fibered with the 
fiber shrinking. The area of the fiber $A$ is 
$$
{M_{Pl}}^3 A = {1 \over R_3}
$$
This can be seen by noting that a membrane wrapped on the fiber 
corresponds to momentum around the circle $R_3$ in the Heterotic 
theory. This limit of M-theory on an elliptically fibered K3 
is exactly what gives F-theory on this K3. The M2-brane probe 
becomes the D3-brane probe in F-theory on K3 \refs{\rSenFO,\rBDS}. Since the 
volume of K3 stays fixed and the fiber shrinks this means that the 
base grows. One might thus think that the moduli space seen by 
the D3-brane probe is infinite. However we should remember 
that a scalar field in 4 dimensions has dimension one, so 
we need a factor of the type-IIB string mass in the area of the 
moduli space.
Inserting this makes the area is up to a constant, $M_s^2$. 
This agrees with the expectation from $S_A(2)$ compactified on 
 $\MT{2}$. We can thus summarize our result for the 4-dimensional 
case. Take the theory $S_A(2)$ with mass scale $M_s$. Compactify it on 
a $\MT{2}$ with R-symmetry twists given by $\tw_1,\tw_2$. 
The $\MT{2}$ is specified by $G^A_{ij},B^A_{ij}$. The moduli space 
of vacua for this $\SUSY{2}$ theory in $D=4$ is the same as the moduli 
space of vacua for the $E_8 \times E_8$ Heterotic 5-brane wrapped 
on $\MT{2}$ with string mass, $M_s$, and a point in 
$O(18,2)/O(18) \times O(2)$ given as follows. The metric 
and 2-form on $\MT{2}$ is $G^A_{ij},B^A_{ij}$. The Wilson lines 
on $\MT{2}$ depend on $\tw_1, \tw_2$. In the case 
$\tw_i=0$ they are the essentially unique Wilson lines 
that break $E_8 \times E_8$ to $SO(8)^4$. For non-zero $\tw_i$ 
the Wilson lines are constructed as in the last section by embedding 
in a diagonal $SU(2)^{16} \subset SO(8)^4$. 

This wrapped 5-brane 
in the Heterotic theory is dual to the 3-brane probe in F-theory on 
the corresponding elliptic-fibered K3.
This K3 is the Seiberg-Witten curve for the 
moduli space. As in the 3 dimensional case, we are not 
saying that the compactified $S_A(2)$ theory is equal to the 
little-string theory on the Heterotic 5-brane, but just that the low-energy 
description is the same. It is obvious that they are not equal 
since the $S_A(2)$ theory has enhanced supersymmetry when $\tw_i=0$.

Decompactifying to 5 dimensions is now easy. The correspondence 
becomes the following. Consider the theory $S_A(2)$ compactified 
on $\MS{1}$ of radius R, string scale $M_s$ and R-symmetry twist $\tw$. 
This is a 5 dimensional theory with $\SUSY{1}$ supersymmetry. The 
coulomb branch is 1-dimensional. Topologically it is $\MS{1}/\BZ_2$. 
This moduli space is the same as the moduli space of the 
Heterotic $E_8 \times E_8$ 5-brane compactified on a circle with 
an $E_8 \times E_8$ Wilson line.
The Wilson line for one $E_8$ is,
$$
W = (0,0,0,0,0,0,0,1)
  + {\tw \over 2\pi} (\half,0,\half,0,\half,0,\half,0),
$$
and the same for the other $E_8$.

Completely analogous statements can be made for the type-IIB 5-brane 
theory, $S_B(2)$. Here the moduli space is given by the 5-brane in the 
Heterotic SO(32) theory. Let us describe this in detail for the case 
of 5 dimensions.
Consider $S_B(2)$ on a circle of radius $R$,
with R-symmetry twist $\tw$ and string scale $M_s$. The moduli space 
of vacua of this is the same as the 5-brane of SO(32) Heterotic 
string theory on a circle with radius R, string scale $M_s$ and SO(32) 
Wilson line
$$
W = (\underbrace{\half,\cdots,\half}_8,\,\underbrace{0,\cdots,0}_8)+
  {\tw \over \pi}(\underbrace{\half,0, \half,0,\cdots,\half,0}_{16})
$$
The string coupling $\lambda$ goes to zero to give a decoupled theory 
on the 5-brane.

There is a dual type-IA picture of the Heterotic theory. The 5-brane 
becomes a D4-brane living on an interval with 8-branes. The parameters 
of the type-IA theory are 
\eqn\etaparam{\eqalign{
M_s' &= {M_s \over \sqrt{\lambda}} \cr
R' &= {\lambda \over {M_s^2 R}} \cr
\lambda' &= {1 \over \sqrt{\lambda}M_s R}
}}
All quantities of interest to the D4-brane theory have a limit as 
$\lambda \rightarrow 0$. The positions of the D8-branes are given 
by the Wilson line.

\fig{The dual type-IA picture.}{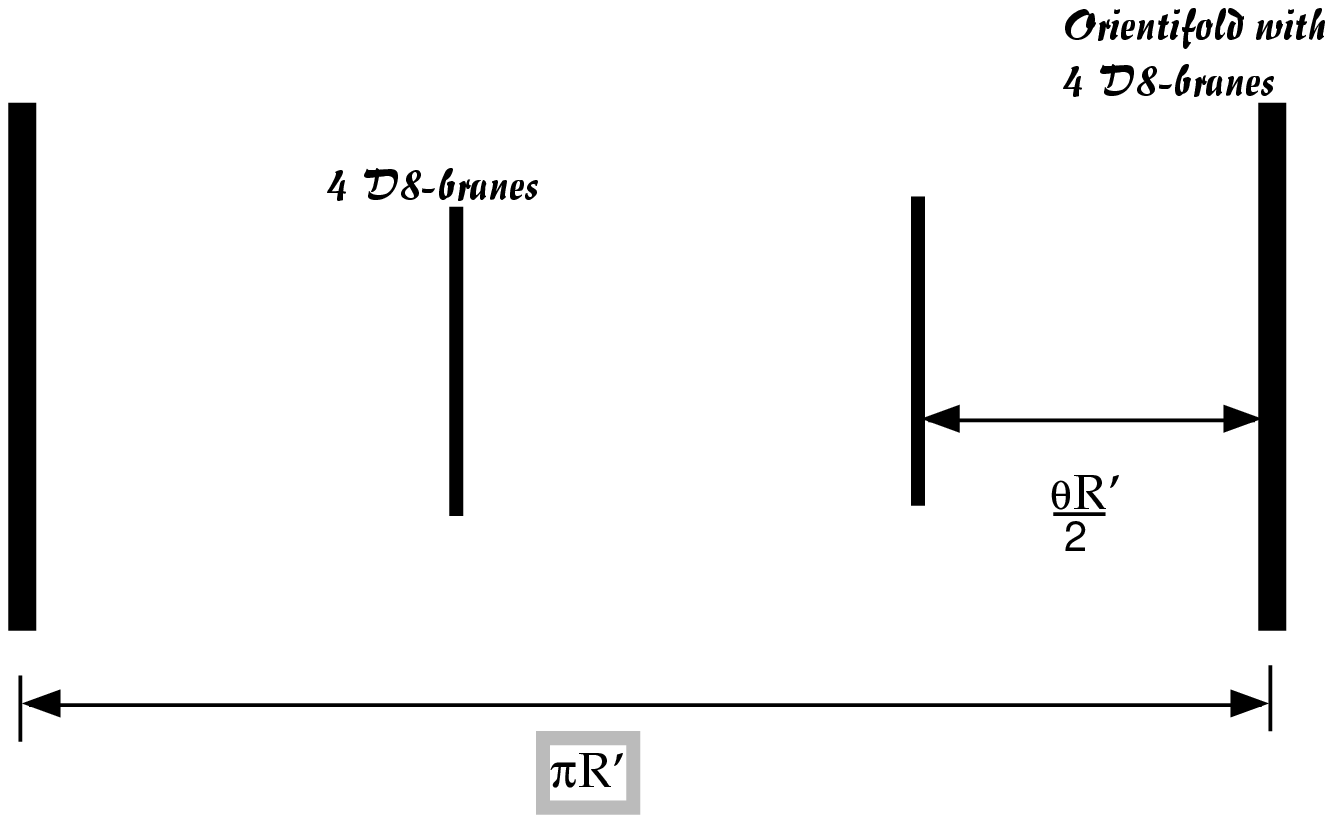} {5truein}

At each end there are 4 D8-branes. 
There are two more stacks of 4 D8-branes at distance ${\tw \over {2}}R'$ 
from each end. When $\tw = 2\pi$ the 8-branes reach 
the other end. This has to be the case since $\tw$ is periodic with 
period $2\pi$.  
We also remark that something interesting happens 
when $\tw = \pi$. Here 8 D8-branes are on top of each other. We will 
return to a discussion of this point later. There behavior in
$D=3,4$ is similar.

$S_A(2)$ compactified to 5 dimensions was described above by
mapping the R-symmetry twists to $E_8\times E_8$ Wilson lines.
For later purposes it will be more convenient to use the type-IA
dual description as the theory living on a D4-brane.
The chain of dualities going from Heterotic 
$E_8 \times E_8$ on $\MS{1}$ to type-IA is to first invoke 
T-duality from Heterotic $E_8 \times E_8$ to Heterotic $SO(32)$, 
and then proceed as above to reach type-IA. The parameters of the 
type-IA theory in terms of the parameters of the $S_A(2)$ theory 
become
\eqn\eotte{\eqalign{
M_s' &= { M_S \over \sqrt{\lambda}}(2({\tw^2 \over \pi^2} + 
          (M_s R)^2))^{1 \over 4} \cr
R' &= {\lam \over M_s^2 R}(2({\tw^2 \over \pi^2} + 
          (M_s R)^2))^{1 \over 2} \cr
\lam' &= {(2({\tw^2 \over \pi^2} + 
          (M_s R)^2))^{5 \over 4} \over \sqrt{\lam}M_s R}
}}
The configuration of D8-branes is as in the $S_B(2)$ case. 
At each end there are 4 D8-branes. At distance ${\tw' \over {2}}R'$ 
from the ends are 4 D8-branes. Here
\eqn\alfaprim{
\tw' = {\tw \over 2({\tw^2 \over \pi^2} + 
          (M_s R)^2)}.
}
We see an interesting effect here. When $\tw =0$, $\tw'$ is also zero. 
For small $\tw$,$\tw'$ is an increasing function. At $\tw = \pi M_s R$, 
$\tw'$ reaches its maximum and start to decrease.

The moduli spaces of all the $S_A(2)$ theories with all possible
$\tw$-twists occupy some subspace 
$$
\Modsp'\subset 
SO(1,17,\BZ)\backslash SO(1,17,\BR)/ SO(17).
$$
We wish to know what is the locus $\Modsp''$
which the $S_A(2)$ theories with the T-dual $\btw$-twists span.
In the above chain of dualities we started with the heterotic 
$E_8\times E_8$ 5-brane wrapped on $\MS{1}$. This represented 
$S_A(2)$ on a circle with a $\tw$-twist.
By definition, $S_A(2)$ with a $\btw$-twist is T-dual to
$S_B(2)$ with some $\tw$-twist and therefore corresponds to
a point in the moduli space of the $SO(32)$ 5-brane.
We have seen that the points on the $SO(32)$ which correspond
to our $S_B(2)$ theories map under heterotic T-duality
to the points on the $E_8\times E_8$ moduli space which correspond
to the $S_A(2)$ theories. Thus $\Modsp'$ and $\Modsp''$ are the
same locus. Nevertheless, $S_A(2)$ with
a $\tw$-twist is not equivalent to $S_A(2)$ with a $\btw$-twist.

The  $S_A(2)$ theory with a $\btw$-twist is T-dual, by definition,
to the $S_B(2)$ theory with an $\tw$-twist. The latter is 
After T-duality to the heterotic $SO(32)$ we would expect

\subsec{A peculiar phase transition} 
As we have explained above, when we compactify $S_B(2)$ with
a twist $\tw$ on $\MS{1}$ of radius $R$ we get a 4+1D theory
whose low-energy is the same as that of a D4-brane probe in
a configuration of D8-branes on an interval. In this configuration
there are 2 stacks of four coincident D8-branes. Whenever 
the D4-brane crosses the stack, a particle of $U(1)$ charge $2$ 
(coming from the adjoint of $SU(2)\supset U(1)$) becomes massless.
When the two stacks of D8-branes coincide we get
two massless hypermultiplets.
Since the low-energy description of a $U(1)$ with two massless
particles is weakly coupled, we can trust field-theory and the
conclusion is that there exists a Higgs phase
where the massless hyper-multiplets get a VEV and break the $U(1)$.

In the $S_B(2)$ case, this phase transition occurs for $\tw=\pi$
and for all values of $R$. In contrast, for $S_A(2)$ this only happens
for small enough $R$. We can see this from eq.\alfaprim.
The phase transition occurs
when $\tw' = \pi$ since this is where two stacks of D8-branes
coincide in the type-IA picture. 
This  has a real solution $\tw$, only if $M_s R \le {1 \over 4}$.
Such a bound is certainly to be expected for $S_A(2)$. The reason 
is that this phase transition happens when two NS 5-branes are 
on opposite points of the 11$^{th}$ circle. The 2 hypermultiplets that 
become massless originate from membranes stretched between the 
two 5-branes and wrapped on the compactified circle (the circle 
which takes us from a 6-dimensional theory to a 5-dimensional theory). 
The tension of the membrane gives a mass to these states. However 
there is also a contribution to the mass from the zero-point energy 
of the fields on the membrane. This contribution depends on $\tw$. 
For certain values of the parameters the zero-point energy can cancel 
the mass from the tension. This is how the hypermultiplet can become 
massless. Obviously the mass from the tension can not be canceled if 
the 11$^{th}$ circle is too big, or equivalently the compactified circle 
is too large. This is the reason for the above inequality.

\subsec{The $(2,0)$ limit}
Let us briefly consider another limit, namely the 
limit where $S_A(2)$ on $\MT{3}$ becomes $T(2)$ on $\MT{3}$. This happens 
when the 11$^{th}$ circle opens up. We see from eq.\volumen\ and eq.\volum\
that $\Vol{K3} \rightarrow \infty$, i.e. the moduli space becomes 
non-compact. This is as expected. Basically we just get half of the K3. 
The other half goes to infinity. In the $E_8 \times E_8$ Heterotic 5-brane
 picture it means that the distance between the ends of the world go to 
infinity and we only look at one end.

\subsec{Field theory limits} 
In this section we will compare the moduli spaces of vacua found in the other
sections with field theory results.  
At each point of the moduli spaces for the $T(k)$ and $S(k)$ theories, 
we can find a field theory description for the light modes. 
We are fortunate that such field theories in $D=3,4,5$ are known.  
The metric on the moduli space around the chosen point will be determined
by the light matter.  We are going to compare our exact metric with this
field theory expectation.  

Let us start with $S_B(2)$ compactified on $\MS{1}$. 
The effective field  theory
is that of the  $D4$-brane probe in type-IA.  
From $S_B(2)$ we have $SU(2)$ gauge theory with (1,1) supersymmetry.
(The field content of the (1,1) vector multiplet are a (1,0) vector multiplet
 and a (0,1) adjoint hypermultiplet.)
Upon compactification on $\MS{1}$ of radius
$R$ with a R-symmetry twist $\tw$, the moduli
space is parameterized by the sixth-component of the gauge field, $A_6$,
 in $U(1)\subset SU(2)$.   
The full R-symmetry of $S_B(2)$ theory is 
$$
SO(4)=SU(2)_U\otimes SU(2)_B
$$
which is broken down to $SU(2)_U$ by $e^{i\tw}\in U(1)\subset SU(2)_B$. 
We get in $5D$ an $SU(2)$ vector-multiplet
and massive adjoint vector-multiplets
(with masses ${n\over {R}}$ with $n\in\BZ_{\neq 0}$)
transforming non-trivially under $SU(2)_U$ R-symmetry.  
The boundary conditions on the two complex scalars in the adjoint
hypermultiplet are shifted by $\tw$:
\eqn\phtwst{
\phi_1(2\pi R) = e^{i\tw}\phi_1(0),\qquad
\phi_2(2\pi R) = e^{-i\tw}\phi_2(0).
}
This shifts the periodicity of the fields around the circle.  
The reduction also gives a tower
of adjoint hypermultiplets in $5D$ with masses,
$$
m^2 = {{(n+ {\tw\over {2\pi}})^2} \over {R^2}}, \qquad n\in \BZ.
$$  
For small $\tw>0$, we get one light adjoint hypermultiplet of mass 
$\tw \over {2 \pi R}$.  Now let us look at the moduli space around $A_6=0$.
From field theory it looks like $SU(2)$ theory with an adjoint hypermultiplet
of mass ${\tw\over {2\pi R}}$.
The gauge coupling is then given by \rSeiFIV,
$$
{1\over {g^2}} = b + c A_6, 
$$  
where  b and c are constants. 
The slope, c, changes when charged matter becomes massless. The change 
in the slope is proportional to the cube of the charge of the 
multiplet becoming massless. In $U(1) \subset SU(2)$ an 
adjoint field has components of charge $-2,0,+2$ under the U(1) in 
units where the $\rep{2}$ of $SU(2)$ has charge $\pm 1$.
This means that the change 
in the slope, c, is 8 times bigger for an adjoint hypermultiplet 
than for a fundamental. Let us calculate at what value of $A_6$ 
the charge 2 component of the adjoint hypermultiplet becomes massless. 
The holonomy around the circle is 
$$
\phi \rightarrow e^{-4\pi i R A_6}  \phi. 
$$
To cancel $e^{i\tw}$ we thus need
$$
A_6 = {\tw \over 4\pi R}.
$$
Let us now compare to the solution from the previous section. Here 
$\tw$ parameterizes the position of 4 D8-branes. For 
$\tw = 2\pi$ they reach the other end of the interval. In terms of 
$A_6$ the other end of the interval is at $1 \over 2R$, so the position 
of the 4 D8-branes is,
$$
A_6= {\tw \over 2\pi}\times {1 \over 2R}= { \tw \over 4\pi R}
$$
in exact agreement.
However the number of D8-branes is 4 and not 8 as naively expected 
from the discussion above. It seems like the change in slope
 is half of what should be expected from field theory. There is 
no discrepancy for a subtle reason. We compare, on one hand, 
 the U(1) low energy 
effective action for a D4-brane moving in an orientifold setting, 
with, on the other hand, a U(1) from a $SU(2)$ gauge theory. The 
U(1) on the D4-brane probe corresponds not to the 
$U(1) \subset SU(2)$ but to one of the U(1) factors in 
$U(1) \times U(1) \subset U(2)$.
The action for the diagonal $U(1)\subset U(2)$ is twice the action
for a single $U(1)$ factor. The normalization would contain
an extra $\sqrt{2}$ factor. Taking this factor of 2 into account
the change in the slope becomes 8 instead of 4.

Let us now consider the case of $S_A(2)$ on $\MT{3}$ with twists 
$\tw_1, \tw_2, \tw_3$. For simplicity the torus is taken 
to be rectangular with radii $R_1,R_2,R_3$ with $B_{ij}=0$. We will 
also take $\tw_i$ to be small.  We want 
to find the light fields. Finding the light fields in this case is 
not as easy as in the previous case, because $S_A(2)$ does not 
have a Lagrangian description. However we can figure out the result 
by first compactifying on a small $R_1$, with $\tw_1 =0$. Then the 
low energy description is a 5-dimensional $\SUSY{2}$ $SU(2)$ gauge 
theory. In $\SUSY{1}$ language it comprises a vector-multiplet and 
a hypermultiplet. Now we can compactify this on a second circle 
of radius $R_2 \gg R_1$. At scale $R_2$ the $SU(2)$ gauge theory is 
weakly coupled and we can perform a classical analysis to include the 
twists $\tw_2$. We get an $SU(2)$ gauge theory in $D=4$ with a 
hypermultiplet of mass $\tw_2 \over 2\pi R_2$. In $D=4$, $\SUSY{2}$ 
a hypermultiplet mass is complex. Since there is no distinction between 
direction 1 and 2 we expect a contribution $\tw_1 \over 2\pi R_1$
from direction 1. They have to combine into a complex mass
$$
m = {\tw_1 \over 2\pi R_1} +i {\tw_2 \over 2\pi R_2}.
$$
On compactifying down to 3 dimensions on $R_3$ (we assume
that $R_3> R_2$) there will similarly 
be a contribution $\tw_3 \over 2\pi R_3$. In $D=3$ a 
hypermultiplet mass consists of 3 real numbers that transform in the 
\rep{3} of $SO(3)$ \rSWGDC.
This $SO(3)$ is part of the R-symmetry group.
We thus 
conclude that the 3 real numbers are $\tw_i \over 2\pi R_i$. There 
is a region in moduli space where the theory looks like $\SUSY{4}$, 
$SU(2)$ gauge theory with an adjoint hypermultiplet with mass 
$m_i =  {{\tw_i} \over {2\pi R_i}}$.
As we have seen, this region is when $|\tw_i|\ll \pi$ and
when the mass scale set by the 2+1D SYM coupling constant (the smallest
of ${{R_1} \over {R_2 R_3}}$, ${{R_2}\over {R_1 R_3}}$ and 
  ${{R_3}\over {R_1 R_2}}$) is much smaller than the smallest
 compactification scale (the smallest of $R_1^{-1}$, $R_2^{-1}$
and $R_3^{-1}$). In our setting, $R_1 \ll R_2 < R_3$, this condition
is met. Note that if $|\tw_i|\ll \pi$ but $R_1\sim R_2\sim R_3$
are of the same order of magnitude, the correct approximation is
to start with the 2+1D CFT to which $\SUSY{8}$ 2+1D SYM flows
\refs{\rSavLen,\rTomNat} and deform it by the relevant operator
to which the mass deformation flows.
When $m_i=0$ we obtain a
$\SUSY{8}$ theory and the moduli space is $(\MR{3} \times \MS{1})/\BZ_2$. 
This has two $A_1$ singularities. When $m_i \neq 0$ these are blown 
up. From our solution in the previous section the sizes of the 
blow up can be read off as a function of $\tw_i$. This means 
that we have derived a formula for the size of the blow-up of 
the singularities in $D=3$, $\SUSY{4}$ $SU(2)$ gauge theory with a 
massive adjoint hypermultiplet.

We can do the same analysis in $D=4$. For $\tw_1 = \tw_2=0$ 
there are 4 singularities. Close to any one of them the system should 
be describable as an $\SUSY{2}$, $SU(2)$ gauge theory with an adjoint 
hypermultiplet. For small $\tw_i$ the mass of the hypermultiplet 
is 
$$
m = {\tw_1 \over 2\pi R_1} +i {\tw_2 \over 2\pi R_2}.
$$
We expect this to change the Seiberg-Witten curve. Our result 
also predicts how this goes. Our picture is that the 
Seiberg-Witten curve is the same as the D3-brane probe in F-theory
on the K3 as described earlier. For $\tw_i=0$ this has a description 
as a type-IIB orientifold 8 plane with 4 D7-branes on top making a 
$D_4$ singularity \refs{\rSenFO, \rBDS}. 
For non-zero $\tw_i$ two of the 7-branes move 
away, giving a $U(2) \times SO(4) = U(2) \times SU(2) \times 
SU(2)$ singularity. In a field theory setting this corresponds to the 
$SU(2)$ Seiberg-Witten theory with 4 fundamental hypermultiplets, 2 of 
them massless and 2 of them massive with equal mass. Our analysis 
thus predicts that this situation should have the same curve as 
the massive adjoint hypermultiplet. In the second Seiberg-Witten paper
\rSWII\ this was indeed found to be the case. In comparing the curves 
with the low energy effective action there is again a factor of 2 in 
the coupling constant $\tau$ because of a
difference in conventions between the adjoint and fundamental case. 
This is the same factor of 2 as explained in the 5-dimensional case above.



\newsec{Reduction of the twisted $(2,0)$ theory to 4+1D}
In this section we will study $T(2)$ on $\MS{1}$ with a twist $\tw$.
Neglecting the overall center-of-mass, the moduli space is
1-dimensional. The low-energy physics is a $U(1)$ vector-multiplet.
Let $\phi$ be the scalar partner of the vector field.
In this section we will study the BPS states in the theory.
There are two different regions in moduli space to consider.
Let $R$ be the radius of $\MS{1}$.
When $\phi R \ll 1$ we can use the effective 4+1D SYM Lagrangian.
We will show that for small $\tw$, the BPS states come from
the $W^\pm$ bosons and the charged states of a massive adjoint 
hyper-multiplet.
When $\phi R \gg 1$ we can identify the charged BPS states with
strings wound around $\MS{1}$.

The BPS masses in 4+1D are \rSeiFIV,
\eqn\bpsmas{
2\phi,\qquad m_0 + 2\phi,\qquad m_0 - 2\phi.
}
In the D4-brane and D8-brane picture, these come from
strings connecting the D4-brane to its image, and to the
two mirror D8-brane stacks. Here,
$$
m_0 = {\tw\over {2\pi R}}.
$$
This can be seen from eq.\eotte\ and eq.\alfaprim.
The states with mass $2\phi$ are vectors while those
with masses $2\phi\pm m_0$ are hyper-multiplets.

\subsec{Yang-Mills limit}
When $\tw=0$, the low-energy description of $T(2)$ on 
$\MS{1}$ is $SU(2)$ SYM with a coupling constant $g^2$ which
is proportional to $R$. As long as our energy scale is 
below the compactification scale $R^{-1}$, the coupling constant
is weak and the effective description is good.
When $|\tw|\ll 1$ it can be incorporated as a small perturbation
in the effective Lagrangian. It corresponds to
giving a bare mass of $m_0$ to the hyper-multiplet in the 
Lagrangian. After spontaneous breaking
of $SU(2)$ down to $U(1)$,
the masses in \bpsmas\ are easily calculated
in field theory. $2\phi$ is the mass of the $W^\pm$ bosons
while $2\phi\pm m_0$ come from the hypermultiplet.
The adjoint hypermultiplet also gives rise to a neutral multiplet
with a mass $m_0$.

\subsec{The large-tension limit}
Let us assume that $\phi R \gg 1$.
In this case, we can first reduce to the 5+1D low-energy of
a single $\SUSY{(2,0)}$ tensor multiplet and then reduce this 
tensor multiplet to 4+1D since the scale of the VEV $\phi$ is
much higher than the compactification scale.
In 4+1D, the neutral states come from the hypermultiplet in
5+1D with twists along $\MS{1}$ as in \phtwst.
The mass of these states is therefore (for small $\tw$),
$$
m = {{\tw}\over {2\pi R}}.
$$
The charged states come from quantization of the strings wrapped
on $\MS{1}$. 
Up to a correction proportional to ${{\a^2}\over {R^2}}$
(see \refs{\rSeiFIV,\rGMS,\rIMS}), the tension of the string
in 5+1D is $\Phi = \phi / 2 R$.
In the limit we are considering, $\Phi R^2\gg 1$, it is enough
to quantize only the low-energy excitations of the strings.
This is just as well, since the low-energy excitations are the only
things we know about these strings!
This means that our results are correct up to $O(1/\Phi R^2)$.
The low-energy description is given by a 1+1D $\SUSY{(4,4)}$ theory.
The VEV of the tensor multiplet of the 5+1D bulk breaks the $Spin(5)$
R-symmetry down to $Spin(4)$.
The 1+1D low-energy description of a string
contains 4 left-moving bosons and 4 right-moving bosons, 4 left-moving
fermions and 4 right-moving fermions.
The bosons are not-charged under the $Spin(5)$ R-symmetry.
The 8 fermions can be decomposed into representations,
of 
$$
(SU(2)_B\times SU(2)_U\times SU(2)'_1\times SU(2)'_2)_{SO(1,1)}
$$
Here $Spin(4)=SU(2)_B\times SU(2)_U$ is the unbroken
R-symmetry of the 5+1D theory,
$Spin(4)=SU(2)'_1\times SU(2)'_2$ is the subgroup of $Spin(5,1)$
of rotations transverse to the string and $SO(1,1)$ is the world-sheet
rotation group.
The fermions are in the 
$$
(\rep{2},\rep{1},\rep{2},\rep{1})_{+\half} +
(\rep{1},\rep{2},\rep{1},\rep{2})_{-\half} 
$$
with an added reality condition.
Under the embedding 
$$
U(1)\subset SU(2)_B\subset SU(2)_B\times SU(2)_U = Spin(4)\subset Spin(5),
$$
we find 2 left-moving fermions with charge $+1$ under $U(1)$,
2 left-moving fermions with charge $-1$ under $U(1)$,
and 4  right-moving fermions with charge $0$ under $U(1)$.
The boundary conditions on the charged fermions are twisted.
Quantization of this system gives low-lying vector-multiplets
and hyper-multiplets with masses,
$$
\Phi R,\qquad
{{\tw}\over {2\pi R}}\pm \Phi R.
$$
Recall that the derivation assumed that
$\Phi R^2 \gg 1$ and $|\tw| \ll \pi$.
This agrees with eq.\bpsmas.



\newsec{R-symmetry twists in the little-string theories}

For the $(2,0)$ theories, which are believed to have a local 
description, a twist by a global symmetry along a circle makes
perfect sense. For the little-string theories, the issue of
locality is more complicated and the meaning of an R-symmetry
twist has to be elaborated.
In this section we will describe the construction in more detail.
We will then see explicitly that T-duality of $S(k)$ does not preserve
the $\tw$-twists. Instead it maps them to T-dual ``$\btw$-twists''.
This raises the intriguing possibility to combine both kinds
of twists simultaneously.

\subsec{Geometrical realization}
One way to define an R-symmetry twist is to realize it geometrically
as follows.
We can start with $\MR{2,1}\times\MR{3}\times\MR{4}$
and mod out by a discrete $\BZ^3$ symmetry which is
generated by elements which act as a shift in $\MR{3}$ and
rotations in $\MR{4}$. We obtain $Z\times\MR{2,1}$
where $Z$ is an $\MR{4}$-fibration over $\MT{3}$.
Explicitly, we define  the 7-dimensional space
$$
Z_{\tw_1,\tw_2,\tw_3} = (\MR{3}\times\MC{2})/\BZ^3,
$$
where $\BZ^3$ is the freely acting group generated by,
\eqn\ygen{\eqalign{
s_1:&
(x_1,x_2,x_3,z_1,z_2)\rightarrow
(x_1 + 2\pi R_1, x_2, x_3,e^{i\tw_1}z_1, e^{-i\tw_1} z_2),\cr
s_2:&
(x_1,x_2,x_3,z_1,z_2)\rightarrow
(x_1, x_2 + 2\pi R_2, x_3, e^{i\tw_2}z_1, e^{-i\tw_2} z_2),\cr
s_3:&
(x_1,x_2,x_3,z_1,z_2)\rightarrow
(x_1, x_2, x_3+2\pi R_3, e^{i\tw_3}z_1, e^{-i\tw_3} z_2),\cr
}}
Here $(x_1,x_2,x_3)$ are coordinates on $\MR{3}$.
We can similarly define 
\eqn\yxdef{
Y_{\tw_1,\tw_2} = (\MR{2}\times\MC{2})/\BZ^2,\qquad
X_\tw = (\BR\times\MC{2})/\BZ.
}
The theory that we study in this paper, $S_A(k)$ on $\MT{3}$ with
a twist, can be obtained if we compactify type-IIA on 
$Z_{\tw_1,\tw_2,\tw_3}$, wrap $k$ NS5-branes on $\MT{3}$ and take
$\lam_s\rightarrow 0$ as in \rSeiVBR.
This shows that it makes sense to include an R-symmetry twist
in $S(k)$.

What is the meaning of these twists in terms of the theory $S(k)$
itself, without appealing to the underlying string-theory?
Let us first refine our terminology.
Let $p$ be a generic point in the parameter space
$$
\Modsp_A \equiv SO(3,3,\BZ)\backslash O(3,3,\BR)/(O(3)\times O(3)).
$$
We will denote the theory derived from $k$ type-IIA NS5-branes at
the type-IIA moduli space point $p\in\Modsp_A$ by $S_A(k;\,p)$.
Similarly there is an identical moduli space $\Modsp_B$ for type-IIB 
NS5-branes. We will denote the theory derived from $k$ type-IIB NS5-branes
at the type-IIB moduli space point $p\in\Modsp_B$ by $S_B(k;\,p)$.
T-duality implies that there is a map,
$$
T:\Modsp_A\rightarrow\Modsp_B,
$$
with $T^2=I$ such that $S_A(k,\,p) = S_B(k;\,T(p))$.
This map can be defined as follows.
Pick an element $v\in O(3,3,\BZ)$ with $\det v = -1$ (all such
elements are $SO(3,3,\BZ)$ conjugate to each other). For
$g\in O(3,3,\BR)$ which is a representative of a point in $p\in\Modsp_A$
take $v\circ g$ to be a representative of $T(p)\in\Modsp_B$.

A generic point $p'$ in the cover,
$$
SL(3,\BZ)\backslash O(3,3,\BR)/(O(3)\times O(3)),
$$
of the parameter space (note that we divided by $SL(3,\BZ)$ instead of
$SO(3,3,\BZ)$) will be called a {\it locality-frame}.
A generic point $p''$ in the cover,
$$
O(3,3,\BR)/(O(3)\times O(3))
$$
will be called a {\it coordinate-frame}.
There are the obvious maps,
$$
p''\rightarrow p'\rightarrow p.
$$
Now suppose that we are in a specific point $p\in\Modsp_A$, say,
and we fix a locality-frame $p'$ for $p$ and a coordinate-frame
$p''$ for $p'$. For a given $p''$ we can contemplate whether it makes
sense to define R-symmetry twists along the cycles of $\MT{3}$.
If they commute with each other,
an $SL(3,\BZ)$ transformation will permute the cycles and will act
on the twists in an obvious way. However, a full
$SO(3,3,\BZ)$ transformation takes one locality-frame to another
and an R-symmetry twist is not mapped back to an R-symmetry twist.

\subsec{The T-dual of an R-symmetry twist}
What does become of an R-symmetry twist after T-duality?
The effect of the R-symmetry twist is to make
a state which is R-charged have a fractional momentum,
because its boundary conditions are not periodic. The
momentum modulo $\BZ$ is related to the R-charge and the twist
in a linear way. Since T-duality replaces the momentum charge
with another $U(1)$ charge -- the winding number of little-strings,
one would deduce that after T-duality, R-charged states should have
fractional winding number.

To be more precise, let us take weakly coupled type-IIA on $X_\tw$
from \yxdef\ and perform T-duality.
Recall that,
$$
X_\tw = (\BR\times\MC{2})/\BZ,
$$
with $\BZ$ generated by,
\eqn\ys{
s: (x,z_1,z_2)\rightarrow (x+2\pi R, e^{i\tw} z_1, e^{-i\tw} z_2).
}
The world-sheet theory is the free type-IIA theory.
Let
\eqn\Xstrmodes{\eqalign{
X =& x+ w\sigma + p\tau 
+ \sum_{n\in\BZ_{\neq 0}} {{\a_{-n}}\over n} e^{i n (\tau-\sigma)}
+ \sum_{n\in\BZ_{\neq 0}} {{\widetilde{\a}_{-n}}\over n}
    e^{i n (\tau+\sigma)},\cr
Z_1 =&
 \sum_{s\in\BZ+\gamma_1} {{\zeta^{(1)}_{-s}}\over s} e^{i s (\tau-\sigma)}
+ \sum_{s\in\BZ+\gamma_1} {{\widetilde{\zeta}^{(1)}_{-s}}\over s}
      e^{i s (\tau+\sigma)},\cr
Z_2 =&
 \sum_{s\in\BZ+\gamma_2} {{\zeta^{(2)}_{-s}}\over s} e^{i s (\tau-\sigma)}
+ \sum_{s\in\BZ+\gamma_2} {{\widetilde{\zeta}^{(2)}_{-s}}\over s}
                  e^{i s (\tau+\sigma)},\cr
}}
$\gamma_{1,2}$ are real numbers which depends on the sector in 
a manner that we will write down below.
When $\gamma_i=0$, we need to add a piece $z_i + p_i \tau$ to $Z_i$.
$p_1,p_2$ are complex while $w,p$ are real.
Also $\a_{-n}^\dagger = \a_n$ and $\widetilde{\a}_{-n}^\dagger 
=\widetilde{\a}_n$.
Let $L$ be the total number of $\zeta^{(1)}$ creation
operators minus the total number of $\zeta^{(2)}$ creation operators
in a state. If some $\gamma_i=0$ we also need to add the rotation
generator $i(z_i p_i^\dagger - z_i^\dagger p_i)$.

\eqn\ndeftot{\eqalign{
L \equiv \sum_{s\in\BZ+\gamma_1} {1\over s}
          (\zeta^{(1)}_{-s})^\dagger \zeta^{(1)}_{-s}
        -\sum_{s\in\BZ+\gamma_2} {1\over s}
          (\zeta^{(2)}_{-s})^\dagger \zeta^{(2)}_{-s}
        + (\zeta \leftrightarrow \widetilde{\zeta}).
}}
Now we can determine which sectors are allowed.
First we require invariance under $s$ in \ys.
This is the world-sheet operator $e^{2\pi i p R - i\tw L}$, so we
require,
$$
p R - {\tw\over {2\pi}} L\in \BZ.
$$
The sector twisted by $s^k$ has
$$
{w\over R} = k,\qquad \gamma = k{{\tw}\over {2\pi}}.
$$

What happens after T-duality?
In a world-sheet formulation, T-duality replaces $p$ with $w$ and
replaces $R$ with $R' = 1/R$. We now have the conditions
$$
{{w'}\over {R'}} - {\tw\over {2\pi}} L \in \BZ,\qquad
p' R' \in \BZ,\qquad \gamma = p' R' {{\tw}\over {2\pi}}.
$$

This suggests a more general twist, which can no longer be described
as modding out by a discrete symmetry.
This time we keep the sectors with
\eqn\abtwists{
p R - {{\tw}\over {2\pi}} L\in \BZ,\qquad
{w\over R} - {{\btw}\over {2\pi}} L\in \BZ,\qquad
2\pi \gamma = \tw {w\over R} + \eta p R.
}
We admit to not having checked that this is consistent with
modular invariance.
The following argument suggests that turning on both  $\tw$ and $\btw$
twists is consistent. For small $\tw$,
turning on a $\tw$-twists corresponds to making a small perturbation
with a certain operator to the Hamiltonian of $S(k)$.
An infinitesimal $\btw$-twist  also corresponds to a perturbation
but with another operator. Now we can make a small perturbation
with both a $\tw$-twist as well as a $\btw$-twist. They preserve
exactly the same supersymmetry. It could, however, happen that after
we turn on both $\tw$-twists and $\btw$-twists there is no longer
any super-symmetric vacuum. We do not know of any way to settle this
question.

We will check in subsequent
sections  what becomes of the \hK\ moduli
spaces after a T-duality.


%
%
\message{S-Tables Macro v1.0, ACS, TAMU (RANHELP@VENUS.TAMU.EDU)}
%
%
\newhelp\stablestylehelp{You must choose a style between 0 and 3.}%
\newhelp\stablelinehelp{You 
should not use special hrules when stretching
a table.}%
\newhelp\stablesmultiplehelp{You have tried to place an S-Table  
inside another
S-Table.  I would recommend not going on.}%
%
%
\newdimen\stablesthinline
\stablesthinline=0.4pt
\newdimen\stablesthickline
\stablesthickline=1pt
%
%
\newif\ifstablesborderthin
\stablesborderthinfalse
\newif\ifstablesinternalthin
\stablesinternalthintrue
\newif\ifstablesomit
\newif\ifstablemode
\newif\ifstablesright
\stablesrightfalse
%
%
\newdimen\stablesbaselineskip
\newdimen\stableslineskip
\newdimen\stableslineskiplimit
%
%
\newcount\stablesmode
\newcount\stableslines
\newcount\stablestemp
\stablestemp=3
\newcount\stablescount
\stablescount=0
\newcount\stableslinet
\stableslinet=0
%
%
%
\newcount\stablestyle
\stablestyle=0
%
%
\def\stablesleft{\quad\hfil}%
\def\stablesright{\hfil\quad}%
%
%
\catcode`\|=\active%
%
%
\newcount\stablestrutsize
\newbox\stablestrutbox
\setbox\stablestrutbox=\hbox{\vrule height10pt depth5pt width0pt}
\def\stablestrut{\relax\ifmmode%
                         \copy\stablestrutbox%
                       \else%
                         \unhcopy\stablestrutbox%
                       \fi}%
%
%
\newdimen\stablesborderwidth
\newdimen\stablesinternalwidth
\newdimen\stablesdummy
\newcount\stablesdummyc
\newif\ifstablesin
\stablesinfalse
%
%
\def\begintable{\stablestart%
  \stablemodetrue%
  \stablesadj%
  \halign%
  \stablesdef}%
\def\stablesadj{%
  \ifcase\stablestyle%
    \hbox to \hsize\bgroup\hss\vbox\bgroup%
  \or%
    \hbox to \hsize\bgroup\vbox\bgroup%
  \or%
    \hbox to \hsize\bgroup\hss\vbox\bgroup%
  \or%
    \hbox\bgroup\vbox\bgroup%
  \else%
    \errhelp=\stablestylehelp%
    \errmessage{Invalid style selected, using default}%
    \hbox to \hsize\bgroup\hss\vbox\bgroup%
  \fi}%
\def\stablesend{\egroup%
  \ifcase\stablestyle%
    \hss\egroup%
  \or%
    \hss\egroup%
  \or%
    \egroup%
  \or%
    \egroup%
  \else%
    \hss\egroup%
  \fi}%
\def\stablestart{%
  \ifstablesin%
    \errhelp=\stablesmultiplehelp%
    \errmessage{An S-Table cannot be placed within an S-Table!}%
  \fi
  \global\stablesintrue%
  \global\advance\stablescount by 1%
  \message{<S-Tables Generating Table \number\stablescount}%
  \begingroup%
  \stablestrutsize=\ht\stablestrutbox%
  \advance\stablestrutsize by \dp\stablestrutbox%
  \ifstablesborderthin%
    \stablesborderwidth=\stablesthinline%
  \else%
    \stablesborderwidth=\stablesthickline%
  \fi%
  \ifstablesinternalthin%
    \stablesinternalwidth=\stablesthinline%
  \else%
    \stablesinternalwidth=\stablesthickline%
  \fi%
  \tabskip=0pt%
  \stablesbaselineskip=\baselineskip%
  \stableslineskip=\lineskip%
  \stableslineskiplimit=\lineskiplimit%
  \offinterlineskip%
  \def\borderrule{\vrule width \stablesborderwidth}%
  \def\internalrule{\vrule width \stablesinternalwidth}%
  \def\thinline{\noalign{\hrule height \stablesthinline}}%
  \def\thickline{\noalign{\hrule height \stablesthickline}}%
  \def\trule{\omit\leaders\hrule height \stablesthinline\hfill}%
  \def\ttrule{\omit\leaders\hrule height \stablesthickline\hfill}%
  \def\tttrule##1{\omit\leaders\hrule height ##1\hfill}%
  \def\stablesel{&\omit\global\stablesmode=0%
    \global\advance\stableslines by 1\borderrule\hfil\cr}%
  \def\el{\stablesel&}%
  \def\elt{\stablesel\thinline&}%
  \def\eltt{\stablesel\thickline&}%
  \def\elttt##1{\stablesel\noalign{\hrule height ##1}&}%
  \def\elspec{&\omit\hfil\borderrule\cr\omit\borderrule&%
              \ifstablemode%
              \else%
                \errhelp=\stablelinehelp%
                \errmessage{Special ruling will not display properly}%
              \fi}%
  \def\stmultispan##1{\mscount=##1 \loop\ifnum\mscount>3
\stspan\repeat}%
  \def\stspan{\span\omit \advance\mscount by -1}%
  \def\multicolumn##1{\omit\multiply\stablestemp by ##1%
     \stmultispan{\stablestemp}%
     \advance\stablesmode by ##1%
     \advance\stablesmode by -1%
     \stablestemp=3}%
  \def\multirow##1{\stablesdummyc=##1\parindent=0pt\setbox0\hbox\bgroup%
    \aftergroup\emultirow\let\temp=}
  \def\emultirow{\setbox1\vbox to\stablesdummyc\stablestrutsize%
    {\hsize\wd0\vfil\box0\vfil}%
    \ht1=\ht\stablestrutbox%
    \dp1=\dp\stablestrutbox%
    \box1}%
  
\def\stpar##1{\vtop\bgroup\hsize ##1%
     \baselineskip=\stablesbaselineskip%
     \lineskip=\stableslineskip%
      
\lineskiplimit=\stableslineskiplimit\bgroup\aftergroup\estpar\let\temp=}%
  \def\estpar{\vskip 6pt\egroup}%
  \def\stparrow##1##2{\stablesdummy=##2%
     \setbox0=\vtop to ##1\stablestrutsize\bgroup%
     \hsize\stablesdummy%
     \baselineskip=\stablesbaselineskip%
     \lineskip=\stableslineskip%
     \lineskiplimit=\stableslineskiplimit%
     \bgroup\vfil\aftergroup\estparrow%
     \let\temp=}%
  \def\estparrow{\vfil\egroup%
     \ht0=\ht\stablestrutbox%
     \dp0=\dp\stablestrutbox%
     \wd0=\stablesdummy%
     \box0}%
  \def|{\global\advance\stablesmode by 1&&&}%
  \def\|{\global\advance\stablesmode by 1&\omit\vrule width 0pt%
         \hfil&&}%
  \def\vt{\global\advance\stablesmode by 1&\omit\vrule width  
\stablesthinline%
          \hfil&&}%
  \def\vtt{\global\advance\stablesmode by 1&\omit\vrule width
\stablesthickline%
          \hfil&&}%
  \def\vttt##1{\global\advance\stablesmode by 1&\omit\vrule width ##1%
          \hfil&&}%
  \def\vtr{\global\advance\stablesmode by 1&\omit\hfil\vrule width%
           \stablesthinline&&}%
  \def\vttr{\global\advance\stablesmode by 1&\omit\hfil\vrule width%
            \stablesthickline&&}%
  \def\vtttr##1{\global\advance\stablesmode by 1&\omit\hfil\vrule  
width ##1&&}%
  \stableslines=0%
  \stablesomitfalse}
\def\stablesdef{\bgroup\stablestrut\borderrule##\tabskip=0pt plus 1fil%
  &\stablesleft##\stablesright%
  &##\ifstablesright\hfill\fi\internalrule\ifstablesright\else\hfill\fi%
  \tabskip 0pt&&##\hfil\tabskip=0pt plus 1fil%
  &\stablesleft##\stablesright%
  &##\ifstablesright\hfill\fi\internalrule\ifstablesright\else\hfill\fi%
  \tabskip=0pt\cr%
  \ifstablesborderthin%
    \thinline%
  \else%
    \thickline%
  \fi&%
}%
\def\endtable{\advance\stableslines by 1\advance\stablesmode by 1%
   \message{- Rows: \number\stableslines, Columns:   
\number\stablesmode>}%
   \stablesel%
   \ifstablesborderthin%
     \thinline%
   \else%
     \thickline%
   \fi%
   \egroup\stablesend%
\endgroup%
\global\stablesinfalse}
%




\newsec{Relation with instantons on non-commutative tori}
So far we have discussed only $T(k)$ and $S(k)$ for $k=2$,
which reduces to $SU(2)$ SYM in appropriate limits.
In this section we are tempted to present conjectures for higher
$k$.
To motivate the conjecture let us first look at the following table,
which lists the limiting moduli spaces when all the R-symmetry twists
are zero, for $S(k)$ on $\MT{3}$, $T(k)$ on $\MT{3}$ and
the masses are zero for
3+1D $SU(k)$ SYM on $\MS{1}$ and 2+1D $SU(k)$ SYM:
\bigskip
\begintable
Theory | $S(k)/\MT{3}$ | $T(k)/\MT{3}$ | 3+1D $SU(k)$ on $\MS{1}$ |
2+1D $SU(k)$ 
\elt
Moduli space |
     $(\MT{4})^k/S_k$ | $(\MT{3}\times\BR)^k/S_k$ |
         $(\MT{2}\times\MR{2})^k/S_k$ | $(\MS{1}\times\MR{3})^k/S_k$
\endtable
\bigskip
As we discussed above
each column is an appropriate limit of the one left to it.
Now let us turn on $\tw$-twists.
As we have argued, in the 3+1D SYM on $\MS{1}$ and
2+1D SYM case, the $\tw$-twists correspond to turning on
a mass to the adjoint hyper-multiplet.
The moduli space of 2+1D $SU(k)$ SYM with
massive adjoint hypermultiplets has been recently
constructed by Kapustin and Sethi \rKS.
It was shown there that this moduli space
is identical to the moduli space of $k$ $U(1)$ instantons on
a non-commutative $\MS{1}\times\MR{3}$.
The moduli space of instantons on non-commutative $\MR{4}$ has
been recently discussed in \refs{\rNS,\rMicha}.
The non-commutativity is characterized by 6 parameters
which transform as a tensor of $SO(4)$. For the moduli space
of instantons one only turns on 3 parameters which transform
as a self-dual 2-form on $\MR{4}$.
In \rKS, the moduli spaces of instantons on the non-commutative 
$\MS{1}\times\MR{3}$ was identified with the moduli space
of the gauge theories by setting the 3 non-commutativity
parameters to be proportional to the 3 mass parameters
of the gauge theory. In fact, Kapustin and Sethi considered
a more general question, namely the moduli space of
3+1D $SU(k)$ SYM with massive adjoint hypermultiplets compactified
on $\MS{1}$. This was mapped to the moduli space of non-commutative
instantons on $\MT{2}\times\MR{2}$.

It is now tempting to conjecture that $S_A(k)$ on $\MT{3}$
with 3 R-symmetry $\tw$-twists has the moduli space of
k $U(1)$ instantons on the non-commutative $\MT{4}$
(obtained from the external parameters as in \slfour).
The non-commutativity is determined by the 3 $\tw_i$'s.

In section (6), we suggested that there is 
a more general perturbation of the $S(k)$ on $\MT{3}$ where
we  turn on both $\tw$-twists and $\btw$-twists.
As we mentioned, it could be that there is no super-symmetric
vacuum after we turn on both twists.
 However, the theory
probably still makes sense. 
Perhaps there is a deeper relation between
non-commutative $\MT{4}$ and the twisted $S(k)$ theories
such that all 6 twists are mapped to all 6 non-commutativity
parameters on $\MT{4}$. The instanton moduli space only depends
on 3 out of the 6 parameters which form the self-dual combination.
Note that in Super-Yang-Mills,
when we turn on a generic configuration
of all 6 non-commutativity parameters, (say in the setting
of \refs{\rCDS,\rDH} with a a D0-brane and $k$ D4-branes 
on a small $\MT{4}$ with some NSNS 2-form flux)
the instanton charge breaks supersymmetry completely.\foot{We are
grateful to M. Berkooz for pointing this out.}
In our setting this might suggest that for the  theories
with a generic configuration of all 6 twists, there is no 
super-symmetric vacuum.



\newsec{A M(atrix) approach}
We will now study the M(atrix) description of the R-symmetry twists.
In the process we will find non-local field theories which
depend on a continuous parameter (the R-symmetry twist)
and which can be mapped to local theories compactified on a smaller space
for rational values of the parameter. This phenomenon is similar
in spirit to Yang-Mills theories on non-commutative spaces as
described in \refs{\rCDS,\rDH,\rCK} and also reminiscent of
the continuous limit of $(p,q)$ 5-branes theories suggested in
\refs{\rWitNGT,\rBarak}.

The M-theory setting that we study is somewhat similar to the
vacua of \rWitNGT\ which were of the form $(\MS{1}\times\MC{2})/\Gamma$
where $\Gamma$ is a discrete subgroup of $SO(4)$ and shifts.

\subsec{The model}
We start with the following geometrical vacuum of M-theory.
Take the 5-dimensional space 
$$
X_\tw = (\BR\times\MC{2})/\BZ
$$
where $\BZ$ is the freely acting group generated by,
\eqn\xmoding{
s: x\rightarrow x+2\pi R,\qquad z_1\rightarrow e^{i\tw} z_1,\qquad
z_2\rightarrow e^{-i\tw} z_2.
}
Here $-\infty < x < \infty $ is a real coordinate on $\BR$ and
$(z_1,z_2)$ are complex coordinates on $\MC{2}$.

For simplicity, let us start with the M(atrix)-model of
M-theory on $X_\tw$ without any 5-branes.
The division by $\Gamma$ is performed according to the rules
of going to the covering space and imposing the following constraints 
\refs{\rBFSS,\rWati}. We pick $U\in U(N)$  and solve,
\eqn\mtxmod{
U X U^{-1} = X + 2\pi R,\qquad
U Z_1 U^{-1} = e^{i\tw} Z_1,\qquad
U Z_2 U^{-1} = e^{-i\tw} Z_2.
}
Here $X$ is the $N\times N$ matrix field corresponding to $x$
in \xmoding\ and $Z_1,Z_2$ are the complex matrices corresponding
to $z_1$ and $z_2$.
Generically, we can assume that $U$ is diagonal and let 
$e^{2\pi i \sigma R}$ be its eigenvalues, with $0\le \sigma\le {1\over R}$.
We assume that each eigenvalue appears $M$ times.
The solution is then, $X = i\px{\sigma} + A(\sigma)$ 
    with $A(\sigma)$ and $M\times M$
gauge field. For $Z_1$ and $Z_2$ we find,
$$
(Z_1)_{\sigma,\sigma'} = 
   \Ph_1({{\sigma+\sigma'}\over 2})\delta(\sigma-\sigma'-{\tw\over {2\pi R}}),
\qquad
(Z_2)_{\sigma,\sigma'} = 
    \Ph_2({{\sigma+\sigma'}\over 2})\delta(\sigma-\sigma'+{\tw\over {2\pi R}}).
$$
Let us define
$$
\xi = {\tw\over {4\pi R}}.
$$
The fields $A(\sigma,\tau)$ and $X(\sigma,\tau)$ are in the adjoint
of $U(N)$ and live on the dual circle of radius $1/2\pi R$.
The fields $\Ph_1(\sigma,\tau)$ 
are not in the adjoint of $U(N)$ but rather
in the product $\rep{N}\otimes \rep{\bar{N}}$
of the gauge group 
$$
U(N)_{(\sigma-\xi)}\otimes U(N)_{(\sigma+\xi)}.
$$
Here $U(N)_{(\sigma)}$ is the group at the point $\sigma$.
In the Lagrangian, to preserve gauge invariance,
the field $\Ph_1(\sigma)$ can be multiplied on the left by local
fields at $\sigma-\xi$ ($X(\sigma-\xi)$ or
$A(\sigma-\xi)$) but on the right, it will have to
be multiplied by local fields as $\sigma+\xi$.
Similarly, $\Ph_1^\dg$ is in $\rep{N}\otimes \rep{\bar{N}}$ of
$U(N)_{(\sigma+\xi)}\otimes
      U(N)_{(\sigma-\xi)}$.
Similar statements hold for $\Ph_2$ and $\Ph_2^\dg$ with
$\sigma+\xi$ replaced by $\sigma-\xi$.
Thus, we will see expressions like
$$
\trp{X(\sigma)\Ph(\sigma+\xi) X(\sigma+2\xi)
 \Ph(\sigma+\xi)^\dg},
\cdots
$$
We denote,
\eqn\covder{\eqalign{
D_\u\Ph_1(\sigma) \equiv& \px{\u}\Ph_1(\sigma) 
    - A_\u(\sigma-\xi)\Ph_1(\sigma)
                 + \Ph_1(\sigma) A_\u(\sigma+\xi),\cr
D_\u\Ph_2(\sigma) \equiv& \px{\u}\Ph_2(\sigma) 
    - A_\u(\sigma+\xi)\Ph_2(\sigma)
                 + \Ph_2(\sigma) A_\u(\sigma-\xi),\cr
D_\u\Ph_1(\sigma)^\dg \equiv& \px{\u}\Ph_1(\sigma)^\dg 
    - A_\u(\sigma+\xi)\Ph_1(\sigma)^\dg
                 + \Ph_1(\sigma)^\dg A_\u(\sigma-\xi),\cr
D_\u\Ph_2(\sigma)^\dg \equiv& \px{\u}\Ph_2(\sigma)^\dg 
   -A_\u(\sigma-\xi)\Ph_2(\sigma)^\dg
                 + \Ph_2(\sigma)^\dg A_\u(\sigma+\xi),\cr
}}
The Lagrangian is given schematically by:
\eqn\mtlag{\eqalign{
{\cal L} =&
{1\over {\lambda}}\int d\sigma d\tau\,\,{\rm Tr}\big\{
D_\u X  D^\u X + F_{\u\v} F^{\u\v} + D_\u\Ph_i D^\u\Ph_i^\dg
\cr &+
\left(X(\sigma)\Ph(\sigma+\xi)
      -\Ph(\sigma+\xi) X(\sigma+2\xi)\right)
\left(X(\sigma+2\xi)\Ph(\sigma+\xi)^\dg
    -\Ph(\sigma+\xi)^\dg X(\sigma)\right)
\cr &+
\left(\Ph_1(\sigma+\xi)\Ph_2(\sigma+\xi)
   -\Ph_2(\sigma-\xi)\Ph_1(\sigma-\xi)\right)
\left(\Ph_1(\sigma-\xi)^\dg\Ph_2(\sigma-\xi)
    -\Ph_2(\sigma+\xi)\Ph_1(\sigma+\xi)\right)
\cr &+
\left(\Ph_1(\sigma+\xi) \Ph_2(\sigma+3\xi)^\dg
-\Ph_2(\sigma+\xi)^\dg \Ph_1(\sigma+ 3\xi)\right)
\cr &\qquad\qquad\qquad
\left(\Ph_1(\sigma-\xi)^\dg\Ph_2(\sigma-3\xi)
   -\Ph_2(\sigma-\xi) \Ph_1(\sigma-3\xi)^\dg\right)
\big\},
\cr
}}
where $\u=0,1$ is along the 1+1D space.
The Lagrangian \mtlag\ is non-local for irrational $\tw$.
For rational $\tw = {m\over l}$ we can redefine the theory on a short
interval of length $1/l$ smaller. We then get a local theory but
with a gauge group $U(N)^l$. The fields $\Ph_i$ become hypermultiplets
in the $(\rep{N},\rep{\bar{N}})$ representation. The matrix model then
reduces to the one described in \rWitNGT, as expected since the
space $X_\tw$ becomes one of the spaces of \rWitNGT\ in this case.

We can now insert $k$ M5-branes at position $Z_1=Z_2=0$.
This amounts to compactifying the model of \rBerDou\ according to
\mtxmod. The new ingredient is that \rBerDou\ also has fields $v$
in the fundamental $\rep{N}$ of $U(N)$. The scalars are not
charged under the $SO(5)$ R-symmetry but they satisfy
$U v = v$ and become localized at impurities at $\sigma=0$ (see 
\refs{\rSav,\rGS,\rKS}).

How do we see the non-trivial moduli space coming out in the M(atrix)
description?
In \rKS\ the Higgs branch of the impurity
system was studied. In these cases, the impurity system was a M(atrix)
model for a system with 16 supersymmetries and the dependence on
external parameters is not expected to be quantum corrected.
In our case, the system is a M(atrix) model for a vacuum with 8
supersymmetries.
One can define a metric on parameter space as,
$$
g_{\a\b}\equiv -\bra{0}\px{\a}\px{\b}\ket{0} - A_\a A_\b,\qquad
A_\a =\bra{0}\px{\a}\ket{0}.
$$
This is very similar to the Zamolodchikov metric in conformal
field-theory.
The relation between this metric and the metric
on the moduli space of the theories will be explored in
a future work \rProg.




\newsec{Discussion}

We have argued that the moduli space of vacua of $S_A(2)$ ($S_B(2)$)
compactified on $\MT{3}$ with 3 R-symmetry twists, $\tw_1,\tw_2,\tw_3$, 
is the same as the moduli space of vacua of the heterotic $E_8\times E_8$
($SO(32)$) $(1,0)$ NS5-brane theory compactified on the same $\MT{3}$
with Wilson lines given by an embedding of the twists in the gauge group.
We have presented a conjecture for higher $k$ involving instantons
on non-commutative tori. 
We have also studied how T-duality of the little-string theory
acts on the R-symmetry $\tw$-twists. We have seen that they get mapped
to other types of twists ($\btw$-twists).
We have suggested that there exist theories with both kinds of twists
simultaneously.

Let us suggest a few questions for further research:
\item{1.}
Confirm or disprove the conjecture of section (7)
about the relation between the moduli spaces of $S(k)$
on $\MT{3}$ and instantons on a non-commutative $\MT{4}$.
\item{2.}
Find an M-theoretic derivation of the moduli spaces, or perhaps
using compactification on a Calabi-Yau manifold.
\item{3.}
Study the BPS spectrum of the theories in 3+1D and 4+1D.
We have identified the moduli spaces of the twisted
$(2,0)$ theory with the moduli space of the compactified 
$E_8$ $(1,0)$ theory. However, these two theories
are not identical. It would be interesting to see
how this distinction is manifested in the multiplicities of
BPS states \refs{\rGTest,\rKMV,\rMNWI,\rMNWII,\rMNVW}.
\item{4.}
Study the M(atrix) models of these compactifications.
In M(atrix)-theory the moduli space of vacua of the theory
should be manifested as the space of external parameters of
the M(atrix) Hamiltonian.
It would be interesting to see how the non-flat metric
and the non-trivial topology of the moduli space arises.
\item{5.}
Study the theories with combined $\tw$-twists and $\btw$-twists.
In particular, do they have a super-symmetric vacuum?
\item{6.} 
Study the other phase where little-strings condense (see section (4.2)).


\bigbreak\bigskip\bigskip
\centerline{\bf Acknowledgments}\nobreak
We wish to thank M. Berkooz, S. Ramgoolam and S. Sethi for discussions.
The research of OJG was supported by a Robert H. Dicke fellowship and by
DOE grant DE-FG02-91ER40671 and the research of
MK was supported by the Danish Research Academy.


\listrefs
\bye
\end